%% file: main.tex
\documentclass[10pt]{article}
\input{packages.tex}

\newif\ifincludesupplement
\includesupplementtrue

\ifincludesupplement
\else
\fi

\title{Synthetic supply networks}
\author[1,2]{Galvin Ng}
\author[3,4]{Luca Mungo}
\author[5]{Damien Bertrand}
\author[3,6]{Fran\c{c}ois Lafond}
\affil[1]{Complexity Science Hub, Vienna, Austria}
\affil[2]{Vienna University of Economics and Business}
\affil[3]{Institute for New Economic Thinking, University of Oxford}
\affil[4]{Macrocosm Inc, New York, USA}
\affil[5]{\'{E}cole Polytechnique F\'{e}d\'{e}rale de Lausanne, Switzerland}
\affil[6]{Smith School of Enterprise and the Environment, University of Oxford}
 
\date{}
\begin{document}
\setlength{\droptitle}{-5em}
\maketitle

\begin{abstract}
\noindent 
A good representation of the population of firms and households is essential for large-scale economic models. While there exist good methods to create synthetic populations of households, creating synthetic populations of firms and, crucially, their supply chain links, is typically much harder. Here, we introduce a flexible method to create synthetic supply networks that match both the known properties of firm-level supply networks and the properties of aggregated input-output tables used in macroeconomic models. Our method is fast, and because it uses only publicly available data, it is fully reproducible and can be easily extended.

\vspace{5mm}
\noindent  Keywords: Synthetic population; Supply networks; Loss optimization
\end{abstract}


\input{sections_for_maindoc/introduction}

\input{sections_for_maindoc/results}
\input{sections_for_maindoc/discussion}

\input{sections_for_maindoc/method}
\input{sections_for_maindoc/end_matters}

\small
\renewcommand\bibname{References}
\bibliographystyle{unsrtnat}
\bibliography{bib}
\normalsize


\ifincludesupplement
\newpage
\appendix
\input{supplementary_body.tex}
\fi

\end{document}

%% file: packages.tex
\usepackage[hmargin=0.9in, vmargin={1in,1in}]{geometry}
\usepackage{amsthm}
\usepackage{tabularx}
\usepackage{amsmath}
\usepackage{amsfonts}
\usepackage{amssymb}
\usepackage{amscd}
\usepackage{mathtools}
\usepackage[dvipsnames]{xcolor}
\usepackage{import}
\usepackage{graphicx}
\usepackage{float}
\usepackage{siunitx}
\usepackage[labelfont=bf]{caption}
\usepackage{csquotes}
\usepackage{kpfonts}
\usepackage{pifont} 

\usepackage{titling} 
\usepackage{authblk}

\usepackage{comment}
\usepackage{wrapfig}
\usepackage{varwidth}

\usepackage{nicefrac}
\usepackage{pdfpages}
\usepackage{multirow}
\usepackage{appendix}
\usepackage{url}
\usepackage{eqparbox}
\usepackage{colortbl}
\usepackage{bbold}
\usepackage{afterpage}
\usepackage{booktabs}
\usepackage{tikz-network}
\usepackage{placeins}
\usepackage[colorlinks]{hyperref}
\hypersetup{
    colorlinks=true,
    allcolors=blue,
    linkcolor=black,
    filecolor=black,      
    urlcolor=black,
    citecolor=Green,
    }
\usepackage[numbers,sort&compress]{natbib}

\usepackage{multicol}

\usepackage{enumitem}
\usepackage{bm}
\usepackage{bbm}
\usepackage{blkarray}
\usepackage{subcaption} 
\usepackage{tikz}
\usepackage{stfloats} 
\usetikzlibrary{positioning,calc,backgrounds}
\usetikzlibrary{datavisualization}
\usetikzlibrary{datavisualization.formats.functions}
\usetikzlibrary{shapes.geometric, arrows}
\tikzstyle{block} = [rectangle, rounded corners, minimum width = 3cm, minimum height = 1cm, text width = 3cm, text centered, draw = black]
\tikzstyle{block_losses} = [rectangle, rounded corners, minimum width = 5cm, minimum height = 1cm, text width = 5cm, text centered, draw = black]
\tikzstyle{block_losses_agg} = [rectangle, rounded corners, minimum width = 3.5cm, minimum height = 1cm, text width = 3.5cm, text centered, draw = black]
\tikzstyle{arrow} = [thick, ->, >= stealth]

\newtheorem{theorem}{Theorem}[section] 
\newtheorem{corollary}{Corollary}[theorem] 
\newtheorem{lemma}[theorem]{Lemma} 

\theoremstyle{remark}

\theoremstyle{definition}

\theoremstyle{definition}

\usepackage[most]{tcolorbox}   
\usepackage[T1]{fontenc}       
\newtcolorbox{blackbox}[1][]{%
  enhanced,
  colback=white,
  colframe=black,
  boxrule=1.8pt,
  arc=8pt,
  left=8pt,right=8pt,top=8pt,bottom=8pt,
  title={#1},
  attach boxed title to top center={yshift=-2mm},
  boxed title style={empty, colback=white},
}

\usepackage{rotating} 

\theoremstyle{plain}
\newtheorem{proposition}[theorem]{Proposition}

\usepackage{threeparttable} 
\usepackage{xr-hyper} 

%% file: sections_for_maindoc/introduction.tex
\section{Introduction}

Economic shocks propagate through transactions between firms, but those transactions are rarely observable. In the few countries in which production networks data is observed, they are highly confidential, and thus hardly usable by the scientific community at large to develop high-resolution economic models. As a result, while there is growing recognition that modeling economic systems at a high level of detail can improve our ability to understand and predict their dynamics~\citep{axtell2022agent,poledna2023economic,wiese2024forecasting,pangallo2024unequal,pangallo2025data,diem2024estimating}, the development of these models faces a serious bottleneck.

This problem is largely solvable by constructing synthetic populations: artificial populations that reproduce key statistical properties of the real system while preserving privacy and filling gaps in the observed data. In fact, for the representation of individual people (as workers, consumers, and members of households), there already exist many methods; the availability of detailed census data has enabled novel computational techniques for generating statistically faithful synthetic populations of households, for instance at the country level \cite{rubinyi2022high,predhumeau2023synthetic,mahmood2025hierarchical,lenti2025population}, global level~\citep{ton2024global,guan2025modeling}, and with applications in policy planning \cite{dyer2024population}, finance \cite{assefa2020generating}, and in healthcare \cite{pangallo2024unequal,giuffre2023harnessing}. 
By contrast, general-purpose, reproducible synthetic populations of firms connected through weighted production networks remain much less developed. Generating a synthetic production network is harder than generating a synthetic population of individuals or households: it is not enough to generate plausible firms in isolation; one must also infer who buys from whom, and along which sectoral and geographic patterns. Aside from aggregate trade flows, which have long been recorded in input-output tables, these properties were, until recently, poorly known.

The situation is, however, beginning to change. Several countries have recently developed complete maps of the business-to-business transactions taking place nationally \cite{pichler2023building}. These datasets have revealed that production networks have statistical properties that are both distinctive and universal \cite{bacilieri2026firm}. This suggests that synthetic supply networks that respect those properties can be very helpful, as they can be close to the real-world network they are substituting for.

Developing synthetic supply networks is essential because, as with synthetic populations of individuals, a crucial issue is that of confidentiality. Statistical agencies and the machine learning community are increasingly concerned about re-identification attacks \citep{rocher2019estimating}, creating an overall push-back on the release of useful micro-data \citep{jordon2022synthetic}. An approach to create synthetic populations of firms would be to train a machine learning model on a confidential dataset, let it learn useful relationships in the data, and use this model to re-create synthetic populations, as suggested by \cite{mungo2023reconstructing}. However, such an approach initially requires access to confidential data. Furthermore, there is always a risk that a model may learn the data ``too closely'', so that releasing its weights would breach privacy to some degree. While the risk appears very low with current methods, the main risk mitigation method (differential privacy) gives guarantees that are hard to interpret \citep{jordon2022synthetic}. Other interesting approaches are being developed, such as federated learning \cite{zheng2023federated} or distributed calibration \cite{garg2024distributed}, which limit the need for complex agreements among data holders but still requires access to confidential data. In any case, the use of confidential data makes results harder to reproduce and extend. As a result, while learning from confidential data to create synthetic networks remains an important avenue of research, here we propose a simpler, more transparent, and immediately useful approach, that can effectively democratize the use of plausible production networks in economic models.

Our approach is entirely reproducible because it \textit{uses only publicly available information}. This is possible thanks to two sources of information: the aforementioned statistics from the micro-data, such as the mean number of customers per firm, and the aggregate input-output tables from the OECD-ICIO dataset \cite{yamano2021development}, which provide payment flows among industries ($45$ industries in our case). Our initial hypotheses was that combining highly distinctive statistics about the structure of the micro-network with meso-level information should help to dramatically restrict the space of plausible networks. While this is likely true, we have also found that several approaches that we have experimented with produce networks that match the targeted statistics, but are nevertheless highly implausible, as judged from visualizing the univariate and bivariate distributions. By contrast, the algorithm we introduce here is able to choose a candidate network that respects the specific statistics used as targets (in a ``soft'' sense), while also resulting in univariate (degrees, strengths, weights) and bivariate distributions that appear reasonable visually. 

Our contribution relates to a rapidly growing literature on supply network reconstruction \cite{mungo2024reconstructing}, which can be roughly divided into two blocks. First, a large body of literature is concerned with individual links. Often motivated by issues of supply chain visibility and supply chain risk management, a large part of the literature has focused on individual links, offering methods to collect data from unstructured sources, infer links using statistical inference methods, and generate synthetic datasets \cite{kosasih2022machine,brintrup2024digital,chang2025learning,vangeneralized,kose2026reconstructing,reisch2022monitoring,mungo2023reconstructing,katafuchi2025construction,long2025leveraging}. 
Second, and thus more related to our goal here, many papers have attempted to construct supply networks with the right system-level properties. This includes a well-established literature leveraging maximum entropy methods \cite{ialongo2022reconstructing, ialongo2024multi, bhattathiripad2025reconstructing, fessina2026inferring, devetak2026industry}, random network formation models \cite{reisch2026supply,bernard2022sparse,henriet2012firm}, or bespoke algorithms \cite{hurt2023supply}.

Our contribution is very distinct from these papers because we clearly state and address a different problem: creating an entirely synthetic population that matches both micro-moments and real-world IOTs, using only publicly and freely available information, and from an algorithm that is transparent, scalable, and can be released open-source. Our synthetic populations can then easily be used to initialize the microstates of $1{:}1$ economic models realistically (as in \cite{henriet2012firm}).  Importantly, the most closely related contributions \cite{hurt2023supply,devetak2026industry,bernard2022origins} use confidential data and/or do not produce networks as close to real-world networks as our method does.

%% file: sections_for_maindoc/results.tex
\section{Results}
\label{sec:results}

\subsection{Reconstruction pipeline and experimental setup}
\label{sub:reconstruction_pipeline_and_setup}

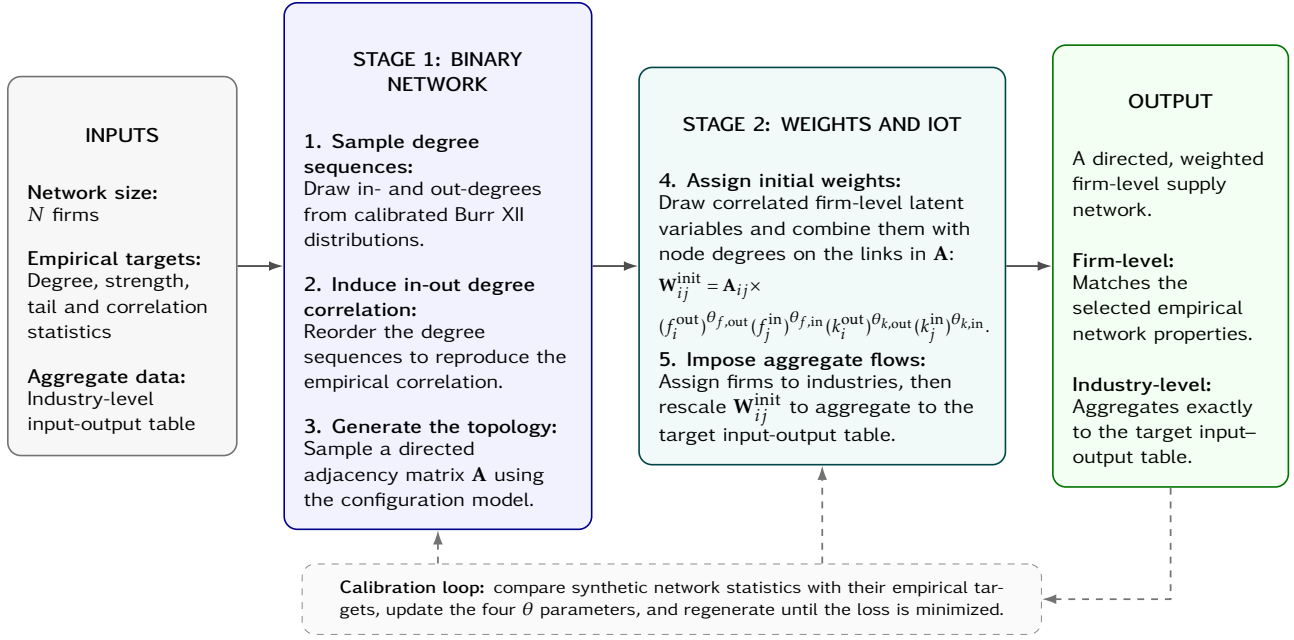
\begin{figure*}[t]
  \centering
  \resizebox{\textwidth}{!}{%
    \input{figures/pipeline}%
  }
  \caption{Overview of the synthetic firm-level supply network algorithm.
  The inputs are the network size, empirical network targets, and a specified industry-level input-output table. The algorithm generates a binary firm-level topology, assigns edge weights, and iteratively calibrates the $\theta$ parameters by comparing the statistics of the generated network with their empirical targets. The resulting directed, weighted firm network reproduces the selected firm-level properties and aggregates exactly to the target input--output table.}
  \label{fig:pipeline}
\end{figure*}

Our method (see \textit{Methods} and the SI for details)  takes as inputs a target number of firms, a set of empirical firm-network properties, and an industry-level input-output table. It then generates a directed, weighted firm-level production network in four main steps (Fig.~\ref{fig:pipeline}). First, we sample separate degree sequences for the number of suppliers and customers, calibrating their marginal distributions to selected empirical properties, including their dispersion and tail behavior. Second, we couple the two degree sequences to reproduce the positive correlation between firms' numbers of suppliers and customers, and use them to generate a directed binary network. Third, we assign preliminary transaction values to the existing links using firm-level latent factors, or \emph{fitnesses},  and node degrees. Finally, we allocate firms to industries and rescale the transaction values within each industry pair so that their aggregation reproduces the target input-output table.

We rely on the findings in \cite{bacilieri2026firm} to determine our reconstruction targets. The study provides a thorough review of all the published information regarding firm-level production networks, as well as an analysis of two datasets based on VAT reporting. It shows that, across different geographies and throughout time, firm-level production networks exhibit some striking statistical regularities. For instance, both in-degree (number of suppliers) and out-degree (number of customers) distributions are fat-tailed; however, the tail exponent of the in-degree distribution is $\sim2.5$, while out-degree distributions typically have much fatter tails, with a tail exponent $\sim 1.5$. Correlations between network properties are also regular: for example, the correlation between expenses and the number of suppliers is lower than the correlation between sales and the number of customers. In the Supplementary Information, we explain how we select which statistics to target, and how we choose their numerical values.

For the main application, we generate a synthetic production network for Hungary in 2015 containing $10^5$ firms. This network size is representative of the country-level production networks surveyed by \cite{bacilieri2026firm}, which contain between $10^4$ and $10^6$ firms. The following section assesses whether the generated network reproduces the targeted firm-level properties, whether its full distributions are empirically plausible, and whether it remains consistent with the aggregate input-output table.

\subsection{Comparison of synthetic and real networks}

\paragraph{Network-level summary statistics.}
Table \ref{tab:main_network_properties} compares the properties of the synthetic network with the corresponding empirical benchmarks, distinguishing between properties that are not targeted and those that are targeted, either when constructing the binary network or when assigning weights.

The generated network closely reproduces the targeted properties of the degree, strength, and edge-weight distributions, as well as the principal relationships between binary and weighted network quantities. While some properties have an almost exact (e.g. tail exponents), others exhibit small deviations (some of the correlations or regressions coefficients). For instance, we reproduce very well the correlation between out-strengths and out-degrees and we slightly underestimate the correlation between in-strength and in-degrees; we still reproduce the fact that strength-degree correlations are smaller on the out- side than on the in-side.

The largest discrepancies are in some features of the binary network, such as reciprocity, clustering, and assortativity. While these are important features of real networks, they were not included in the calibration targets, and, in general, can't be reproduced with the configuration model that we used to generate the binary network. In fact, the configuration model does not include mechanisms that explicitly promote reciprocal relationships, triadic closure, or assortative matching between firms \cite{newman2018networks}, but we keep it here as it is simple and very fast.

\input{tables/table1.tex}

\paragraph{Visual evidence on distributional properties.}

\begin{figure}[!htbp]
    \centering
    \includegraphics[width=0.97\linewidth,keepaspectratio]
    {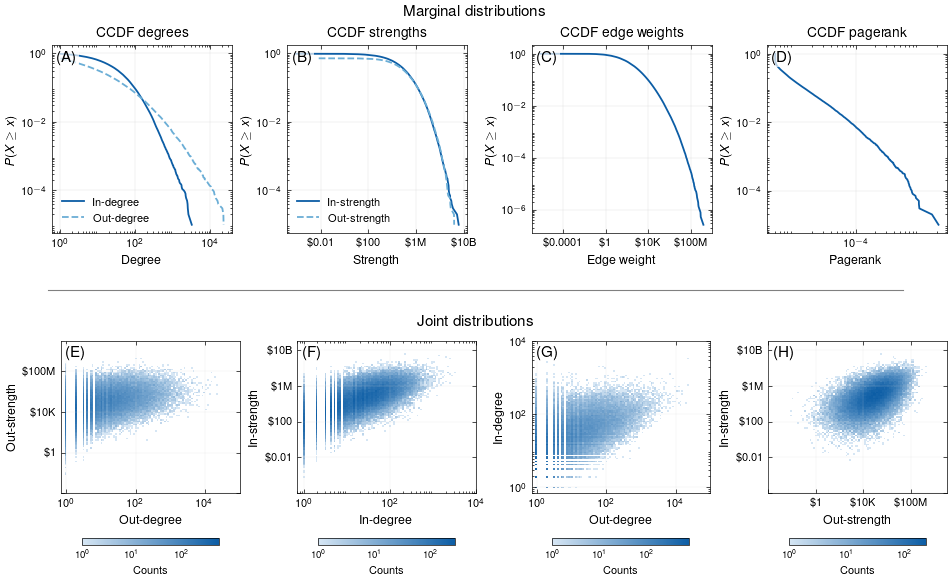}
    \caption{Distributional properties of a generated synthetic supply network of $100,000$ firms, aggregated to the 2015 Hungarian input-output table. (A-D) Complementary cumulative distribution functions (CCDFs) of key node and edge-level variables. Strengths and weights are measured in USD. (E-H) Joint distributions (two-dimensional histograms with logarithmic color scaling) of key pairs of variables. All axes are displayed in logarithmic scales.}
    \label{fig:main_network_visualization}
    \vfill
    \vspace{4mm}
    \includegraphics[width=0.85\linewidth, keepaspectratio]
    {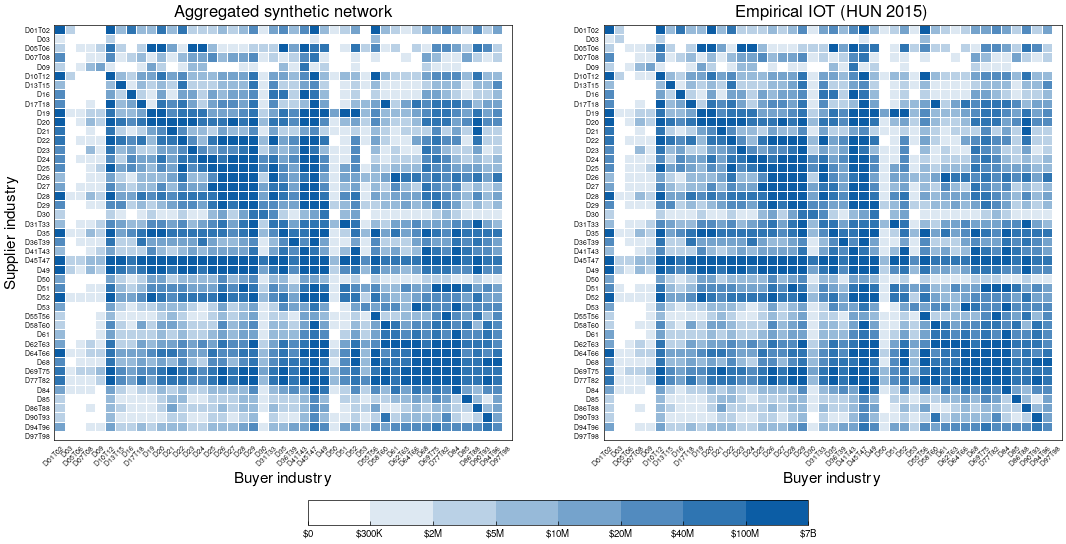}
    \caption{Comparison between the aggregated synthetic firm-level production network and the empirical 2015 Hungarian input-output table (IOT). The left panel shows the industry-level transactions obtained by aggregating the synthetic firm-level network, while the right panel shows the empirical Hungarian 2015 IOT. Colors correspond to common quantile bins computed jointly across both matrices, allowing direct comparison of transaction flows. The synthetic network reproduces the industry flows observed in the empirical IOT.}
    \label{fig:main_iot_visualization}
\end{figure}

Our experience creating networks that match the properties in Table \ref{tab:main_network_properties} suggests that some procedures and optimization algorithms may achieve what seem good results according to these statistics, yet deliver networks that are clearly pathological when the distributions are visualized. For this reason, we show in Figure \ref{fig:main_network_visualization} the univariate and joint distributions generated by the model.

The synthetic network reproduces the asymmetric degree distributions and symmetric strengths distributions, showing patterns that closely resemble those published in Ref~\cite{bacilieri2026firm} and in the references cited therein  (note, however, that for the joint distributions, Ref. \cite{bacilieri2026firm} obfuscates bins with less than 10 observations for confidentiality reasons, so our figures are not perfectly comparable).

\paragraph{Consistency with IO tables.}

Finally, Fig.~\ref{fig:main_iot_visualization} compares the targeted Input-Output matrix with the one obtained by aggregating our firm-level synthetic network, showing an excellent agreement. The agreement is expected because the rescaling procedure explicitly enforces consistency with the target input-output table. However, it is not trivial that rescaling micro-values to ensure meso-scale properties keeps micro-scale properties unaffected.

%% file: figures/pipeline.tex
\begin{tikzpicture}[
  font=\sffamily\small,
  >={Latex[length=2.2mm,width=1.5mm]},
  flow/.style={->, line width=0.8pt, draw=black!70},
  feedback/.style={->, dashed, line width=0.75pt, draw=black!55},
  base/.style={
    rounded corners=2mm,
    line width=0.7pt,
    minimum height=58mm,
    inner sep=3mm,
    align=left,
    execute at begin node={\hyphenpenalty=10000\relax}
  },
  inputbox/.style={base, text width=29mm, draw=black!55, fill=black!3},
  stageone/.style={base, text width=41mm, draw=blue!55!black, fill=blue!5},
  stagetwo/.style={base, text width=50mm, draw=teal!55!black, fill=teal!5},
  outputbox/.style={base, text width=30mm, draw=green!45!black, fill=green!5!white},
  loopbox/.style={
    rounded corners=2mm,
    draw=black!45,
    dashed,
    fill=black!2,
    text width=108mm,
    inner sep=2.5mm,
    align=center,
    font=\sffamily\footnotesize
  }
]

\node[inputbox] (inputs) {
  \begin{center}\textbf{INPUTS}\end{center}
  \vspace{2mm}
  \textbf{Network size:}\\[-0.5mm]
  $N$ firms

  \vspace{3mm}
  \textbf{Empirical targets:}\\[-0.5mm]
  Degree, strength, tail and correlation statistics

  \vspace{3mm}
  \textbf{Aggregate data:}\\[-0.5mm]
  Industry-level \\ input-output table
};

\node[stageone, right=7mm of inputs] (topology) {
  \begin{center}\textbf{STAGE 1: BINARY NETWORK}\end{center}
  \vspace{2mm}
  \textbf{1. Sample degree sequences:}\\[-0.5mm]
  Draw in- and out-degrees from calibrated Burr XII distributions.

  \vspace{3mm}
  \textbf{2. Induce in-out degree correlation:}\\[-0.5mm]
  Reorder the degree sequences to reproduce the empirical correlation.

  \vspace{3mm}
  \textbf{3. Generate the topology:}\\[-0.5mm]
  Sample a directed adjacency matrix $\mathbf{A}$ using the configuration model.
};

\node[stagetwo, right=7mm of topology] (weights) {
  \begin{center}\textbf{STAGE 2: WEIGHTS AND IOT}\end{center}
  \vspace{2mm}
  \textbf{4. Assign initial weights:}\\[-0.5mm]
  Draw correlated firm-level latent variables and combine them with node degrees on the links in $\mathbf{A}$:
  \\[1mm]
    {\footnotesize
    \(
    \begin{aligned}
    &\mathbf{W}_{ij}^{\mathrm{init}} = \mathbf{A}_{ij} \times \\
    &(f_i^{\mathrm{out}})^{\theta_{f,\mathrm{out}}}
    (f_j^{\mathrm{in}})^{\theta_{f,\mathrm{in}}}
    (k_i^{\mathrm{out}})^{\theta_{k,\mathrm{out}}}
    (k_j^{\mathrm{in}})^{\theta_{k,\mathrm{in}}}.
    \end{aligned}
    \)
    }
    \\[1mm]
  \textbf{5. Impose aggregate flows:}\\[-0.5mm]
  Assign firms to industries, then rescale $\mathbf{W}_{ij}^{\mathrm{init}}$ to aggregate to the target input-output table.
};

\node[outputbox, right=7mm of weights] (output) {
  \begin{center}\textbf{OUTPUT}\end{center}
  \vspace{2mm}
  A directed, weighted firm-level supply network.

  \vspace{4mm}
  \textbf{Firm-level:}\\[-0.5mm]
  Matches the selected empirical network properties.

  \vspace{4mm}
  \textbf{Industry-level:}\\[-0.5mm]
  Aggregates exactly to the target input--output table.
};

\draw[flow] (inputs.east) -- (topology.west);
\draw[flow] (topology.east) -- (weights.west);
\draw[flow] (weights.east) -- (output.west);

\node[loopbox, below=15mm of weights, xshift=-23mm] (calibration) {
  \textbf{Calibration loop:}
  compare synthetic network statistics with their empirical targets,
  update the four $\theta$ parameters, and regenerate until the loss is minimized.
};

\draw[feedback] (output.south) |- (calibration.east);
\draw[feedback] (topology.south |- calibration.north) -- (topology.south);
\draw[feedback] (weights.south |- calibration.north) -- (weights.south);

\end{tikzpicture}

%% file: tables/table1.tex
\begin{table}[t]
\centering
\small

\setlength{\heavyrulewidth}{1.3pt}
\setlength{\lightrulewidth}{0.8pt}
\setlength{\cmidrulewidth}{0.8pt}

\begin{threeparttable}

\begin{tabular}{@{}c@{\hspace{0.04\textwidth}}c@{}}
\toprule

\begin{tabular}[t]{@{}lrrc@{}}
Metric & Synthetic & Empirical & Target \\
\cmidrule(lr){1-4}
Number of nodes & 100,000 & 100,000 & B \\
Number of edges & 3,825,747 & 4,000,000 & B \\
Share of $k^{\mathrm{in}}=0$ & 0 &  & NT \\
Share of $k^{\mathrm{out}}=0$ & 26.4 &  & NT \\
Mean degree & 38.3 & 40.0 & B \\
\cmidrule(lr){1-4}
Max $k^{\mathrm{in}}$ & 3,190 &  & B \\
Max $k^{\mathrm{out}}$ & 22,462 &  & B \\
Mean log $k^{\mathrm{in}}$ & 3.28 & 3.0 & NT \\
Var log $k^{\mathrm{in}}$ & 1.31 & 2.0 & B \\
Mean log $k^{\mathrm{out}}$ & 2.18 & 2.0 & NT \\
Var log $k^{\mathrm{out}}$ & 2.95 & 3.0 & B \\
\cmidrule(lr){1-4}
Mean log $s^{\mathrm{in}}$ & 10.8 & 10.0 & NT \\
Var log $s^{\mathrm{in}}$ & 8.88 & 9.0 & W \\
Mean log $s^{\mathrm{out}}$ & 10.9 & 10.0 & NT \\
Var log $s^{\mathrm{out}}$ & 9.18 & 8.0 & W \\
\cmidrule(lr){1-4}
LWCC & 95,556 &  & NT \\
Avg path length & 2.71 & 3.0 & NT \\
Degree assortativity & -0.0688 & [-0.2, -0.01] & NT \\
Reciprocity & 0.00671 & [0.03, 0.05] & NT \\
Avg clustering & 0.0987 & [0.19, 0.28] & NT \\
Global clustering & 0.0208 & low & NT \\
\end{tabular}

&

\begin{tabular}[t]{@{}lrrc@{}}
Metric & Synthetic & Empirical & Target \\
\cmidrule(lr){1-4}
Hill $\alpha$, $k^{\mathrm{in}}$ & 2.53 & 2.5 & B \\
Hill $\alpha$, $k^{\mathrm{out}}$ & 1.46 & 1.5 & B \\
Hill $\alpha$, $s^{\mathrm{in}}$ & 1.1 & 1.0 & W \\
Hill $\alpha$, $s^{\mathrm{out}}$ & 1.09 & 1.0 & W \\
Hill $\alpha$, influence & 1.45 & [1.2, 1.3] & NT \\
Hill $\alpha$, weights & 1.09 & [1.1, 1.2] & NT \\
\cmidrule(lr){1-4}
Corr: $k^{\mathrm{in}}\sim k^{\mathrm{out}}$ & 0.552 & 0.55 & B \\
Corr: $s^{\mathrm{in}}\sim s^{\mathrm{out}}$ & 0.505 & 0.5 & NT \\
Corr: $s^{\mathrm{out}}\sim k^{\mathrm{out}}$ & 0.442 & 0.5 & W \\
Corr: $s^{\mathrm{in}}\sim k^{\mathrm{in}}$ & 0.592 & 0.75 & W \\
\cmidrule(lr){1-4}
OLS: $s^{\mathrm{out}}\sim k^{\mathrm{out}}$ & 0.779 & 0.76 & W \\
OLS: $k^{\mathrm{out}}\sim s^{\mathrm{out}}$ & 0.251 & 0.33 & W \\
OLS: $s^{\mathrm{in}}\sim k^{\mathrm{in}}$ & 1.54 & 1.4 & W \\
OLS: $k^{\mathrm{in}}\sim s^{\mathrm{in}}$ & 0.227 & 0.4 & W \\
\cmidrule(lr){1-4}
TLS: $s^{\mathrm{in}}\sim s^{\mathrm{out}}$ & 0.968 & 1.0 & W \\
TLS: $k^{\mathrm{in}}\sim k^{\mathrm{out}}$ & 0.496 & 0.7 & NT \\
\cmidrule(lr){1-4}
IOT RMSE (100 MUSD) & 8.01e-04 &  & W \\
\end{tabular}

\\
\bottomrule
\end{tabular}

\begin{tablenotes}
\footnotesize
\item[B] Targeted when creating adjacency matrix $\mathbf{A}_{ij}$.
\item[W] Targeted when creating weights $W_{ij}$ on existing links $\mathbf{A}_{ij}$.
\item[NT] Not targeted.
\end{tablenotes}

\end{threeparttable}

\caption{Network properties of a generated synthetic supply network with $100{,}000$ firms, aggregated to the 2015 Hungarian input-output table. Values are reported to $3$ significant figures, except for the number of nodes and edges, which are reported as integers.}
\label{tab:main_network_properties}
\end{table}

%% file: sections_for_maindoc/discussion.tex
\section{Discussion}

Our method provides plausible synthetic networks, compatible with macro-economic aggregates, without having direct access to administrative data. This paves the way for a straightforward research agenda developing synthetic populations for all networks underlying the functioning of economic systems as measured in national accounts, pursuing an agenda advanced at a more aggregate level by \cite{andersen2026disaggregated} (see also \cite{buda2023national} for a bottom-up reconstruction of national accounting objects in the case of consumption). This would include bipartite networks linking workers to the firms that employ them, and consumers to the firms they buy from. This can be achieved using a research pipeline similar to the one we introduced here, after researchers with access to scanner data, transaction and/or survey data (for consumer-to-firm networks, and employee-employer networks (for firm-to-worker networks) make key moments available publicly. Similarly, the study of loan-level data could provide key moments to create realistic representation of the real-financial nexus of the economy. 

Our approach is deliberately simple, leaving room for increased complexity when more targetable statistics will become available. As international institutions develop distributed micro-data projects, more and more useful moments of firm networks will become available, including inter-country trade networks, geographic patterns, more detailed insights about industry-level patterns, or using the recently developed reconstructions of product-product networks \cite{fetzer2024ai,fessina2025product,karbevska2025mapping,o2025deciphering} instead of the very coarse IOTs we have been using. New topological features can be incorporated as an explicit step in the construction of the adjacency matrix. For instance, while preserving the degree distribution, one can scramble the pairing of existing edges to generate the desired levels of reciprocity and clustering in the synthetic network. Features involving edge weights could potentially be incorporated by extending the loss function. It is also possible to extend our method for use when the identity of the firms and their (intermediate) sales are known, although there would likely be an incompatibility between observed firm sales and IOTs, so one would have to decide which one should be preserved.

This point raises a notable conceptual limitation of our work. As explained in some details in \cite{bacilieri2026firm}, IOTs do not record transactions in the same way as firm-level VAT data does, because a number of accounting conventions differ, for instance the treatment of wholesale and retail, transport margins, or the financial sector. We know from \cite{bacilieri2026firm} that this problem is somewhat limited outside of specific sectors, so here we have ignored it. A simple partial solution would be to leverage the recent and important work in Ref. \cite{colon2025constructing}, who reintroduce wholesale and retail trade flows in IOTs, making them much closer conceptually to VAT datasets.

Our approach is straightforwardly testable. We know that some countries do have VAT data, but have not published anything about it yet. Our paper makes very clear testable predictions: one can generate synthetic networks for these countries now, and compare the synthetic networks to the actual networks only in the future. An interesting test would be to compare real and synthetic networks not only in terms of topological and statistical features, but also in terms of the behavior of specific models or stress-test algorithms such as the ESRI \cite{diem2022quantifying}, as suggested in \cite{mungo2024reconstructing}.

Our paper also provides a new problem for the machine learning community, with benchmark results to be improved upon. Our pipeline does not use generative AI methods. We have found that existing methods in AI-based graph generation did not meet our graph needs or could not be adapted in ways that were effective. We offer a clear challenge to the computer science community: develop AI-based graph generation methods that beat our established benchmark while remaining tractable for graphs at the national scale ($10^6$). An interesting development would be to leverage the work on synthetic populations of LLM-based personas \cite{park2023generativeagents,ge2024personahub,tang2026llmsynthor,paglieri2026personagenerators} to create behaviorally plausible synthetic populations of firms operating in a supply chain. Our work provides a benchmark for the topological structure, and we think a useful method to initialize such approaches.

Our code is fully available and implementable in a differentiable framework, so potentially re-usable by novel approaches in that space. Our algorithm is available open source, with an easy implementation for all countries covered by the OECD ICIO tables. The user only needs to decide on a country to target and a number of firms. A configuration file allows users to modify the target values for the elements of the loss.

%% file: sections_for_maindoc/method.tex
\small
\section*{Methods}
We generate the synthetic firm-level supply networks in two phases: first, creating a directed binary network consisting of firm-to-firm connections; second, populate edge weights conditional on the existing links targeting the correlation structure between strengths and degrees, and the aggregated economic flows from the industry-level input-output table. We provide a brief summary here. See the SI a full explanation, and the codebase for all the implementation details. 

\subsection*{Generating the binary network}
We first draw out- and then in-degree sequences from Burr XII distributions, with strictly positive parameters $(c,k,\lambda)$. The Burr XII distribution $X \sim \text{BurrXII}(c,k,\lambda)$ is characterized by three parameters $c$, $k$, $\lambda$. Its cumulative distribution function (CDF) is given by 
$F(x) = 1 - \left[ 1 + \left( x/\lambda \right)^{c} \right]^{-k},$
where the asymptotic tail exponent is
$\alpha = ck$,\textbf{} since $1-F(x) \sim x^{-ck}$ as $x \to \infty$.
We sample degrees from the Burr XII distribution to control the tail exponent, mean, and log-variance of the degree distributions. We calibrate the parameters numerically so that the final distribution matches user-defined targets. Since the Burr XII distribution is continuous and supported on $(0,\infty)$, the sampled observations must subsequently be discretized, truncated, and reconciled so that the total in-degree equals the total out-degree. 

Because the firms' in- and out-degree sequences are drawn independently from Burr XII distributions, their correlation is initially zero. To reproduce the empirical log in- and out-degree correlation of approximately $0.55$ reported by \cite{bacilieri2026firm}, we first rank firms according to their in-degrees and perturb the ordering using multiplicative lognormal noise with rank-dependent variance,
$
k^{\text{in, scrambled}}_i = k^{\text{in}}_i \times \varepsilon_i$, with $\varepsilon_i \sim \text{Lognormal}\left(-\frac{1}{2}(1-\tilde{r}_i)^{2\zeta}, (1-\tilde{r}_i)^{2\zeta}\right)$
where $\tilde{r}_i \in [0,1]$ denotes the normalized rank of firm $i$'s in-degree.
The out-degree sequence is then reordered to match the ranking of the scrambled in-degrees.
This procedure preserves the marginal degree distributions while inducing a tunable positive correlation between in- and out-degrees. We calibrate $\zeta$ to match the empirical Pearson correlation reported by \cite{bacilieri2026firm}, obtaining $\zeta=-0.14$.

We then sample an adjacency matrix $\mathbf{A}$ from the configuration model with the in-degree and reordered out-degree sequences as inputs, and remove parallel edges and self-loops.

\subsection*{Generating the weighted network}

We then assign weights to the edges in the adjacency matrix $\mathbf{A}$. We do this by first generating ``fitnesses'' (latent variables for each node, a classic idea in network science), and then using these and node degrees in a gravity-like formula to determine edge-level weights,
\begin{equation}
\mathbf{W}_{ij}^{\text{init}} = (f_{i}^{\text{out}})^{\theta_{f,\mathrm{out}}}(f_{j}^{\text{in}})^{\theta_{f,\mathrm{in}}} (k^{\text{out}}_i)^{\theta_{k,\mathrm{out}}} (k^{\text{in}}_j)^{\theta_{k,\mathrm{in}}}\mathbf{A}_{ij},
\label{eq:initial_weights_maindoc}
\end{equation}
where $f$ are the fitnesses and $\boldsymbol{\theta} \equiv (\theta_{k,\mathrm{in}} \;,\;\theta_{k,\mathrm{out}}\;,\;\theta_{f,\mathrm{in}}\;,\;\theta_{f,\mathrm{out}})$ denotes a set of tunable parameters. We then rescale $\mathbf{W}_{ij}^{\text{init}}$ to match the input-output table (IOT). We first describe how we determine the fitnesses and how we rescale to match the IOT, assuming the parameters $\boldsymbol{\theta}$ are known. Next, we describe our procedure to find optimal values of $\boldsymbol{\theta}$.

\paragraph{Assignment of weights assuming $\boldsymbol{\theta}$ is known}

We sample in- and out-fitnesses from a joint lognormal $(\ln f^{\text{in}}, \ln f^{\text{out}}) \allowbreak \sim \mathcal{N}( \boldsymbol{\mu}_f, \boldsymbol{\Sigma})$. While the choice of the mean does not affect the result, we choose the covariance matrix carefully. Note that in Eq. \ref{eq:initial_weights_maindoc}, we assign each fitnes variable an exponent $\theta$, which we optimize. Since the variance of the variable $f^\theta$ can be tuned arbitrarily by changing the value $\theta$, we choose the variance of $\ln f$ in a way that we facilitate parameter search, by ensuring that all four $\theta$s are likely to be on similar scales. Intuitively, we expect the four $\theta$s to be on similar scales if the four variables are also on similar scales. Therefore, we assume that the variances of the log in- and log out-fitnesses are equal to the variances of the log in- and log out-degrees. Finally, for the covariance, we introduce a parameter $\rho$ that tunes the correlations, and which we calibrate to $0.8$, ensuring that the resulting log in- and log out-strength correlation is close to the empirical value of $0.5$.

To match the IOT given edge-level weights $\mathbf{W}_{ij}^{\mathrm{init}}$, we assume that the number of firms per industry is proportional to industry sales. 
Firm ordering follows industry ordering: the first $M_1$ firms are assigned to industry 1, the next $M_2$ firms to industry 2, and so on. While proportional allocation appears as a reasonable assumption, the method is compatible with any desired allocation strategy. With firms assigned to industries, we rescale the firm-to-firm weights to ensure that the industry-to-industry weights match a given IOT, using
\begin{equation*}
\mathbf{W}_{ij} = \frac{\text{IOT}_{{g_i}{g_j}}}{\sum\limits_{f \in g_i}\sum\limits_{h \in g_j} \mathbf{W}_{fh}^{\text{init}}} \mathbf{W}_{ij}^{\text{init}}.
\end{equation*}

\paragraph{Estimation of the optimal parameters $\boldsymbol{\theta}$}

Finally, we calibrate the parameters
$\boldsymbol{\theta}
=(\theta_{k,\mathrm{in}},\theta_{k,\mathrm{out}},
\theta_{f,\mathrm{in}},\theta_{f,\mathrm{out}})$
so that the generated network reproduces selected marginal and joint
properties of firms' strengths and degrees. Using Optuna
\cite{akiba2019optuna}, we minimize

\begin{equation*}
\mathcal{L}
=
\mathcal{L}_{\mathrm{tail}}
+\mathcal{L}_{\mathrm{corr}}
+\mathcal{L}_{\mathrm{OLS}}
+\mathcal{L}_{\mathrm{TLS}}
+\lambda_{\mathrm{var}}\mathcal{L}_{\mathrm{var}}.
\end{equation*}
For any statistic $q$, let $\mathcal{D}(q)=\left[\widehat{q}(\boldsymbol{\theta})-q^\star\right]^2$ denote the squared discrepancy between its value in the synthetic network,
$\widehat{q}(\boldsymbol{\theta})$, and its empirical target, $q^\star$.
The tail component is
$\mathcal{L}_{\mathrm{tail}}=\mathcal{D}(\alpha^{\mathrm{in}})+\mathcal{D}(\alpha^{\mathrm{out}}),$
where $\alpha^{\mathrm{in}}$ and $\alpha^{\mathrm{out}}$ are the tail
exponents of the in- and out-strength distributions, estimated using the
Hill estimator.
The correlation component is
$\mathcal{L}_{\mathrm{corr}}=\mathcal{D}(\rho^{\mathrm{in}})+\mathcal{D}(\rho^{\mathrm{out}}),$
where $\rho^{\mathrm{in}}$ is the Pearson correlation between log
in-strength and log in-degree, and $\rho^{\mathrm{out}}$ is the corresponding correlation between log out-strength and log out-degree. The OLS component is
$
\mathcal{L}_{\mathrm{OLS}}
=
\mathcal{L}^{\mathrm{in}}_{\mathrm{OLS}}
+
\mathcal{L}^{\mathrm{out}}_{\mathrm{OLS}},
$
with
$
\mathcal{L}^{\mathrm{in}}_{\mathrm{OLS}}
=
\mathcal{D}(\beta^{\mathrm{in}}_{s\leftarrow k})
+
\mathcal{D}(\beta^{\mathrm{in}}_{k\leftarrow s})
$
and
$
\mathcal{L}^{\mathrm{out}}_{\mathrm{OLS}}
=
\mathcal{D}(\beta^{\mathrm{out}}_{s\leftarrow k})
+
\mathcal{D}(\beta^{\mathrm{out}}_{k\leftarrow s}).
$
Here, $\beta^{\mathrm{in}}_{s\leftarrow k}$ is the OLS slope obtained by
regressing log in-strength on log in-degree, whereas
$\beta^{\mathrm{in}}_{k\leftarrow s}$ is obtained from the reverse
regression. The out-strength coefficients are defined analogously. The arrow
points from the explanatory variable to the dependent variable. The TLS component is
$
\mathcal{L}_{\mathrm{TLS}}=\mathcal{D}(\tau^{\mathrm{in}\leftarrow\mathrm{out}}),
$
where $\tau^{\mathrm{in}\leftarrow\mathrm{out}}$ is the total least squares
slope relating log out-strength to log in-strength. Finally, the variance component is
$
\mathcal{L}_{\mathrm{var}}=\mathcal{D}(v^{\mathrm{in}})+\mathcal{D}(v^{\mathrm{out}}),
$
where $v^{\mathrm{in}}$ and $v^{\mathrm{out}}$ are the variances of log
in-strength and log out-strength, respectively. All loss components have unit
weight except for the variance component, which is multiplied by
$\lambda_{\mathrm{var}}=10^{-3}$. All correlations, regression coefficients,
and variances are computed using log-transformed variables.

In practice, we repeat the Optuna search procedure $200$ times, with each search consisting of $200$ trials.
Across these repeated searches, we observe stable optimal parameter regions: the optimal value of $\theta_{k,\mathrm{in}}$ is centered around $0$, while $\theta_{k,\mathrm{out}}$ is consistently negative. On the other hand, $\theta_{f,\mathrm{in}}$ and $\theta_{f,\mathrm{out}}$ are either both negative or positive. 
Additional details on the parameter search are provided in the Supplementary Information, where we also show that networks generated using negative $\theta_{f,\mathrm{in}}$ and $\theta_{f,\mathrm{out}}$ are statistically indistinguishable from the positive case. 
For the remainder of the paper, we use the parameters
$(\theta_{k,\mathrm{in}} \;,\;\theta_{k,\mathrm{out}}\;,\;\theta_{f,\mathrm{in}}\;,\;\theta_{f,\mathrm{out}}) = (0, -0.7, -1.3, -1.3)$.

Randomness enters the methodology through Burr degree draws, the reordering procedure for out-degrees, sampling the adjacency matrix from the configuration model, sampling of fitnesses, and firm-to-industry assignment. 
Our methodology is implemented using PyTorch sparse COO tensors, so computational and memory requirements scale with the number of edges rather than the size of the full matrix. 
Since the generated networks have approximately constant mean degree, this corresponds to $O(N)$ rather than $O(N^2)$ scaling.
We generate synthetic supply networks of sizes up to $500,000$ firms, while preserving firm-level network properties and exact aggregation with input-output tables. 
This scalability makes the framework suitable for large-scale agent-based and macroeconomic applications.

\normalsize

%% file: sections_for_maindoc/end_matters.tex
\subsection*{Code availability}
The code that has been used to obtain all results in the paper is publicly available on \url{https://github.com/galvinngkw/synthetic_supply_networks}.

\subsection*{Acknowledgments}
G.N. was supported by the Digital Innovation School, which is funded by the Federal Ministry of Women, Science and Research (BMFWF) and the Federal Chancellery (BKA) of Austria. 
F.L. would like to thank Baillie Gifford, UK Research and Innovation through Economic and Social Research Council grant PRINZ (ES/W010356/1), and the Institute for New Economic Thinking at the Oxford Martin School for funding. F.L. is an affiliated scientist with Macrocosm Inc. 
We are grateful to Anton Pichler, Jan Hurt, and audiences at CSH, WIFO, Centro Enrico Fermi, IMT Lucca, SMU Trade Conference 2026, APCNCS 2026, and CEF 2026 for many useful comments and discussions.

\subsection*{Author contributions}
F.L., G.N and L.M. conceived the work and developed the methodology. 
G.N. and L.M. wrote the code.
G.N. performed the data analysis. 
G.N., F.L. and L.M. analyzed and interpreted the results. 
All of the authors contributed to the final manuscript.

\subsection*{Competing interests}
The authors declare no competing interests.

%% file: supplementary_body.tex
\setlength{\heavyrulewidth}{1.3pt}
\setlength{\lightrulewidth}{0.8pt}
\setlength{\cmidrulewidth}{0.8pt}

\begin{center}
\section*{Supplementary Information}
\end{center}

In this SI, we (1) discuss how we choose which statistics to target, (2) explain our approach in full detail, both for the binary network creation step, and the weight attribution step, and (3) we provide robustness checks.

Our full initialization procedure is detailed in the following subsections, and is outlined in Box \ref{box:generation_procedure_optuna}.

\begin{figure}[!htbp]
\centering
\input{tables/generation_procedure_in_box}
\captionsetup{name=Box}  
    \caption{Procedure to generate a synthetic network. Refer to the individual subsections for details.}
\label{box:generation_procedure_optuna}
\end{figure}

\section{Choosing target values}
\label{subsection:choosing_target_values}

We choose which statistics are useful to target, and their numerical values, by a careful analysis of the results in \cite{bacilieri2026firm}, who found that a number of network statistics are very similar across countries (see especially their summary in Table 9). Ref. \cite{bacilieri2026firm} synthesizes the literature, and analyzes two ``complete'' networks (based on VAT, no reporting thresholds) themselves. However, the number of networks for which statistics are reported remains low, so there is some arbitrariness in determining target values and acceptable ranges for the statistics. Similarly, Ref. \cite{bacilieri2026firm} reports basic statistics, to allow for a synthesis of a larger literature, but does not report more sophisticated network properties. 
Here we are limited by this, but in principle other statistics could be added as targets in our optimization algorithm (although it would likely need to be adapted).

Table \ref{tab:choosing_empirical_values} provides details of how we select the target values used.

\begin{figure}[!htbp]
    \centering
    \includegraphics[width=\linewidth]{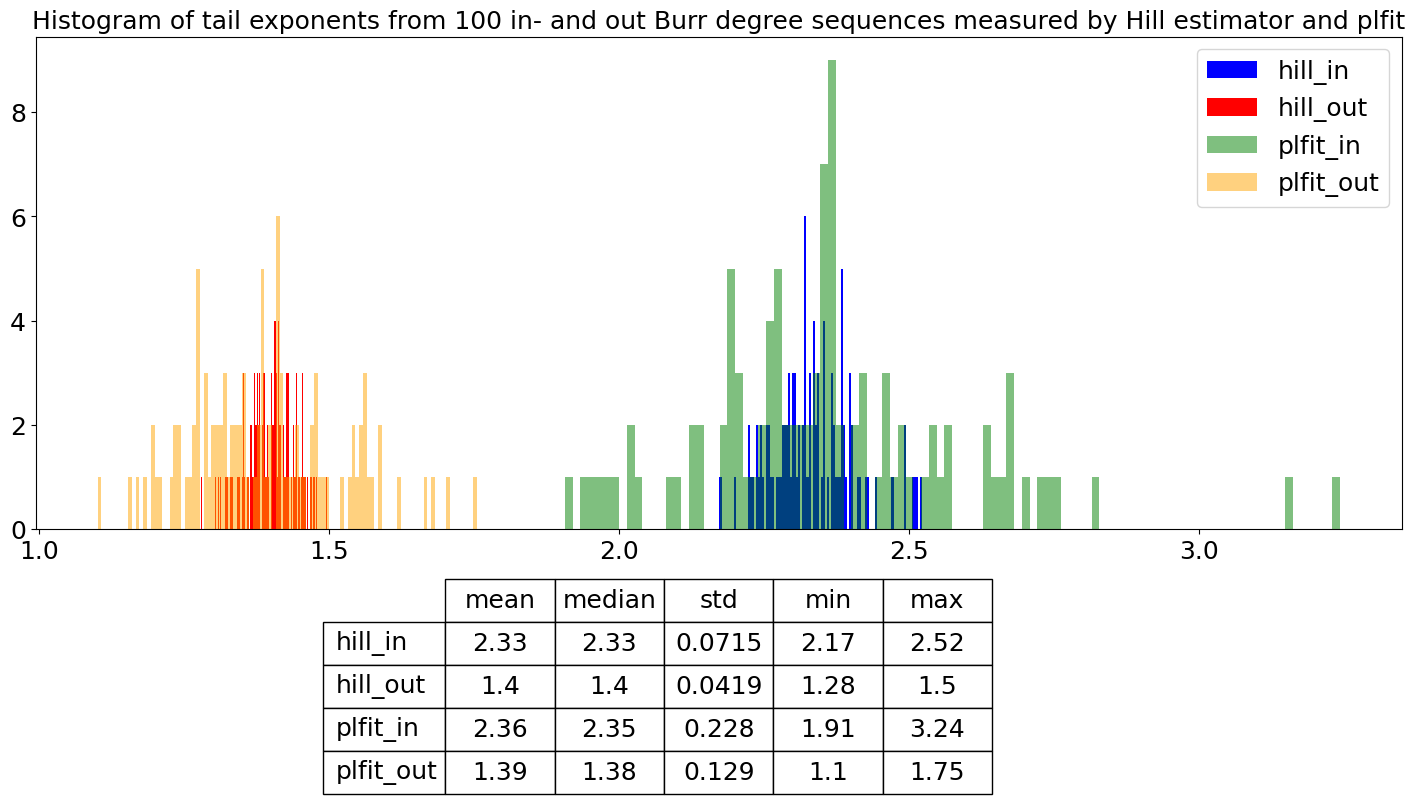}
    \caption{We generate $100$ independent pairs of in- and out degree sequences using $(\alpha,\nu) = (2,10)$ for the out-degrees and $(\alpha,\nu)=(3,3)$ for the in-degrees. For each pair, tail exponents are estimated using the Hill estimator based on the top $1\%$ of degrees and compared with estimates obtained from igraph \texttt{plfit}. The Hill and \texttt{plfit} estimators have nearly identical means, though \texttt{plfit} shows greater variability.}
    \label{fig:hill_deg_threshold_percentile_against_plfit}
\end{figure}

For power law tail exponents, a threshold above which the tail is measured should be determined. \citet{bacilieri2026firm} reports the results from various estimators that compute this threshold endogenously, showing some variation (especially for in-degrees, as expected because the exponent is higher), and report the Newman-Clauset-Shalizi \cite{clauset2009power} Hill estimator (\texttt{plfit}) in their main text. For optimization purposes, it is easier to specify the threshold in advance. We have chosen a threshold  roughly in line with the threshold found endogenously by \texttt{plfit} in \cite{bacilieri2026firm}. Taking the example of degree distributions, we have chosen a threshold of $1\%$ (see Table \ref{tab:choosing_empirical_values}). Figure~\ref{fig:hill_deg_threshold_percentile_against_plfit} confirms that our procedure generates networks with not only the right exponent when measured using a fixed 1\% threshold, but also when measured using the \texttt{plfit} algorithm. The exponent measured using the \texttt{plfit} algorithm is substantially more volatile across simulation runs \-- which is why our algorithm implements a fixed threshold.

\input{tables/choosing_target_val}

\FloatBarrier
\newpage
\section{Generating firm-level binary networks}
\label{appendix_generating_binary_networks}
\subsection{Generate out-degrees sequences}
\label{subsubsection:generate_burr_out_deg}

We first draw out- and then in-degree sequences from Burr XII distributions, with strictly positive parameters $(c,k,\lambda)$. Denote a Burr XII random variable by 
\begin{equation}
\label{eq:burr_eq}
X \sim \text{BurrXII}(c,k,\lambda),
\end{equation}
with its Cumulative Distribution Function (CDF) given by 
\begin{equation} 
F(x) = 1 - \left[ 1 + \left( \frac{x}{\lambda} \right)^{c} \right]^{-k}.
\label{eq:burr_cdf}
\end{equation}
Let $U$ be a Uniform$[0,1]$ random variable. Then,
\begin{equation} 
X = \lambda \left(  \left(1 -U \right)^{-\frac{1}{k}} - 1 \right)^{\frac{1}{c}}.
\label{eq:burr_from_uniform_rv}
\end{equation}

We want to find parameters of the Burr XII distribution that will give observations with specific tail exponent, mean and variance of the log-transformed values. From the CDF in Eq. \eqref{eq:burr_cdf}, it is clear that as $x \to \infty$, $1-F(x) \sim x^{-ck}$, so the asymptotic tail exponent is
$$
\alpha = ck.
$$

Then for a user-defined tail exponent $\alpha > 1$, mean $E[X] \equiv \mu$, and log-variance $\text{Var} (\ln X) \equiv \nu$ we can find the Burr XII parameters $c,k,\lambda$ as follows. We solve $\nu = f(\alpha, k)$ numerically for $k$, where $f$ is  
\begin{equation}
\nu = \frac{ \frac{\pi^2}{6} + \psi_1(k)}{\left( \alpha/k \right)^2},    
\label{eq:nu_formula}
\end{equation}
and $\psi_1(\cdot)$ is the trigamma function (See Corollary \ref{corollary:soln_for_mean_and_logvar} for its proof). Knowing $k$, the parameters $c$ and $\lambda$ are then determined using
\begin{equation}
    c = \frac{\alpha}{k},
    \quad 
    \lambda = \frac{\mu}{kB\left(k - \frac{1}{c}, 1 +\frac{1}{c}\right)},
\label{eq:c_and_lambda_formula}
\end{equation}
where $B(\cdot,\cdot)$ is the Beta function. In principle, this gives us a straightforward mapping, helping us choose parameters of the Burr XII distribution that will give us the desired measured statistics.

However, in practice, we cannot obtain a distribution with the desired properties simply by sampling from a Burr XII distribution with these analytically determined statistics, for three reasons: (1) the tail exponent described above is valid only asymptotically, (2) the Burr XII distribution is continuous, so we have to discretize it, and (3) we have to truncate the distribution because firms can not have more  partners than the number of other firms. We describe next each of these issues, and then how they impact naive Burr XII draws, and how we have dealt with this in our final algorithm.

\paragraph{Finite size tail exponent.} While asymptotically, the tail of the Burr XII distribution behaves as a power law with slope $\alpha = c k$, at each finite percentile the slope can be quite different. In Corollary \ref{corollary:log-log_slope_of_burr}, we calculate how the slope of the CCDF changes over the percentiles. We find that the complementary cumulative distribution function (CCDF), defined as $\bar{F}(x) = 1 - F(x)$, has a $\log$-$\log$ slope given by
\begin{equation}
-\frac{d\log \bar F(x)}{d\log x} = 
\alpha \left( 1 - p^{\frac{1}{k}}\right) = k \sqrt{\frac{\frac{\pi^2}{6} + \psi_1(k)}{\nu}} \left(1 - p^{\frac{1}{k}}\right),
\label{eq:log_log_ccdf_slope}
\end{equation}
evaluated at the point $x_p$ satisfying $\bar{F}(x_p) = p$. The value $p$ denotes the tail probability at which the slope is calculated. Equation \eqref{eq:log_log_ccdf_slope} implies that the $\log$-$\log$ slope depends on both tail probability\footnote{
See Section \ref{subsection:choosing_target_values} for a discussion of the choice of tail probability $p$ used in estimating the Hill exponents.
} $p$ and log-variance $\nu$.

\paragraph{Discretization.} Since the Burr XII distribution is continuous, we discretize the sampled values so that they represent integer-valued firm degrees. We do this by rounding to the nearest integer.

\paragraph{Truncation.} Although the Burr XII distribution is supported on $(0,\infty)$, the theoretical maximum degree of a firm in a firm-level network is the number of firms. Moreover, empirical evidence in \cite{bacilieri2026firm} shows that roughly, the largest out-degree is about an order of magnitude smaller than the number of nodes, and the largest in-degree is about an order of magnitude smaller than the largest out-degree. To capture these two properties in our synthetic binary network, we discard and resample any sampled out-degrees greater than $40\%$ of the total number of firms, and any sampled in-degree greater than $5\%$ of the number of firms. 

\paragraph{Consequences for the choice of parameters.} For the tail exponent, we have seen from Eq. \eqref{eq:log_log_ccdf_slope} that the estimated tail exponent depends on the tail probability $p$ and the log-variance $\nu$. To better understand how discretization and truncation affect the final measured statistics of draws from a Burr XII distribution, we have performed an ablation study. Table \ref{tab:integerize_truncate_reconcile_degrees_hill001} shows the results. Discretizing Burr XII observations lowers the realized log-variance relative to the initial value $\nu$. As shown in the bottom-most block, for $N=100,000$, an out-degree sequence drawn from the Burr XII distribution with $\nu=10$ has a log-variance of $10$, as expected; after discretization, however, the measured value decreases to $2.95$. Truncating the discretized observations has negligible effect on both the tail exponent and log-variance, aside from reducing the maximum out-degree from $85,028$ to $30,616$. 

To solve these issues, we calibrate $\alpha$ and $\nu$ by trial and error, and then compute the corresponding Burr XII parameters $(c,k,\lambda)$ using Equations \eqref{eq:nu_formula} and \eqref{eq:c_and_lambda_formula}.

\begin{figure}[!htbp]
    \centering
    \includegraphics[width=\linewidth]{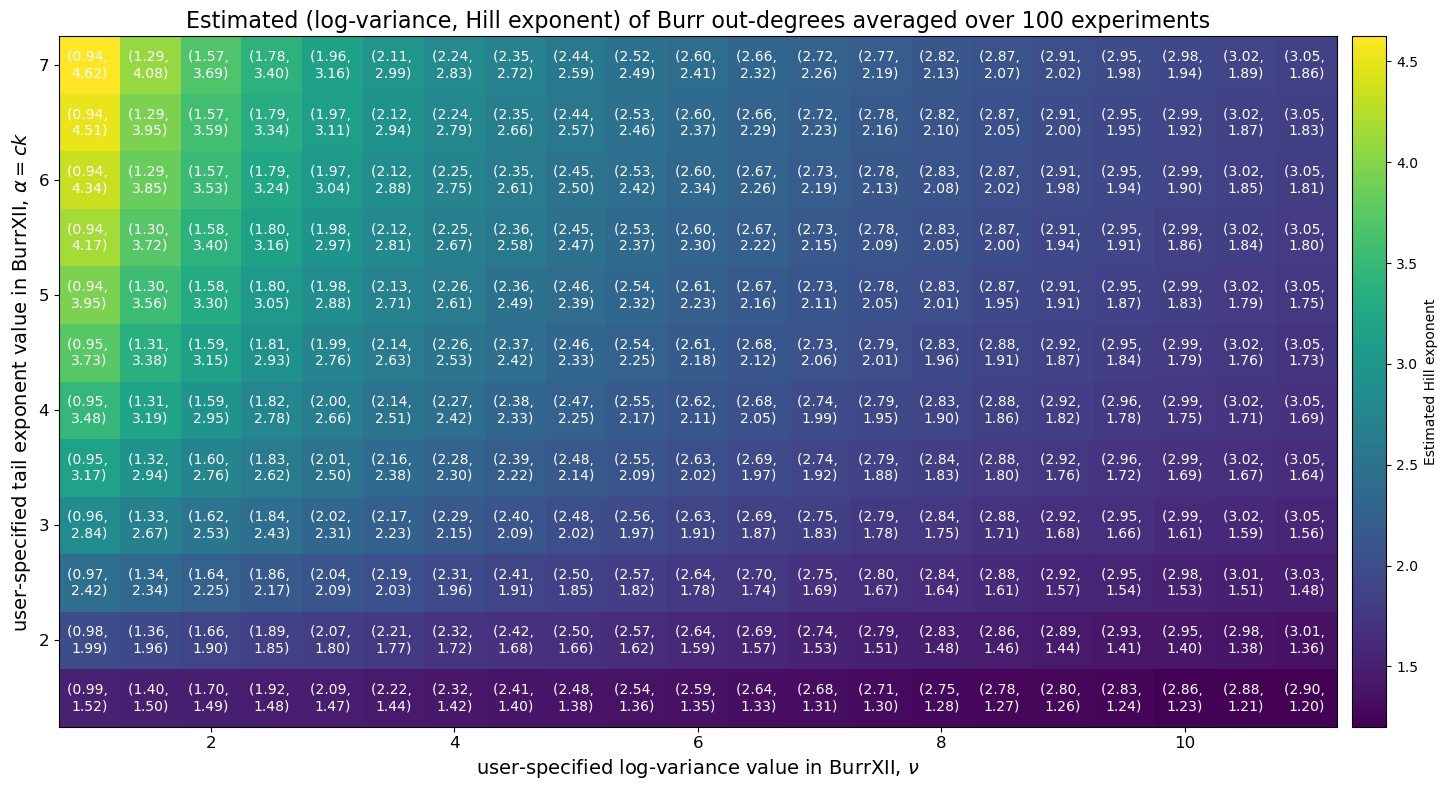}
    \caption{Meshgrid with step size $0.5$ over input tail exponent $\alpha$ and log-variance $\nu$ for out-degrees. For each grid point, we generate $5,000$ independent sequences of size $100,000$. The realized Hill exponents and log-variances are displayed in white. We select $(\alpha, \nu) = (2,10)$ as input values, as they yield a realized Hill exponent of $1.4$ and log-variance of $2.95$. We deliberately select an $\alpha$ that produces a realized Hill exponent below the empirical value of $1.5$, since sampling adjacency matrices via the Configuration Model induces an upward bias, increasing the final Hill exponent in the synthetic network from $1.4$ to $1.5$. Since the Configuration Model does not affect the log-variance, $\nu$ is chosen to match the empirical value of $3.0$ as closely as possible.}
    \label{fig:burr_alpha_nu_out_degree_meshgrid}
\end{figure}

Figure \ref{fig:burr_alpha_nu_out_degree_meshgrid} presents a grid over input tail exponent $\alpha$ and log-variance $\nu$ for out-degrees. To generate a truncated out-degree sequence for a network of $N=100,000$ firms, mean $\mu \approx 40$, log-variance $\approx 3$, tail exponent $\approx 1.4$, and a maximum out-degree capped at $0.4N$, this meshgrid shows that values of $\alpha = 2$ and $\nu = 10$ will produce the desired properties.

\input{tables/ablation_study}

Finally, to ensure robustness against stochastic variation in truncated Burr XII draws, for networks with $N=100,000$ firms, we retain only out-degree sequences with tail exponent in the range $1.4 < \alpha < 1.5$, log-variance $\nu > 2.95$, and a maximum out-degree of at least $10\%$ of firms.
This restriction guarantees that the largest out-degree lies between $10\%$ and $40\%$ of the network size, consistent with empirical magnitudes.

\subsection{Generate Burr in-degrees}
\label{subsubsection:generate_burr_in_deg}
\begin{figure}[!htbp]
    \centering
    \includegraphics[width=\linewidth]{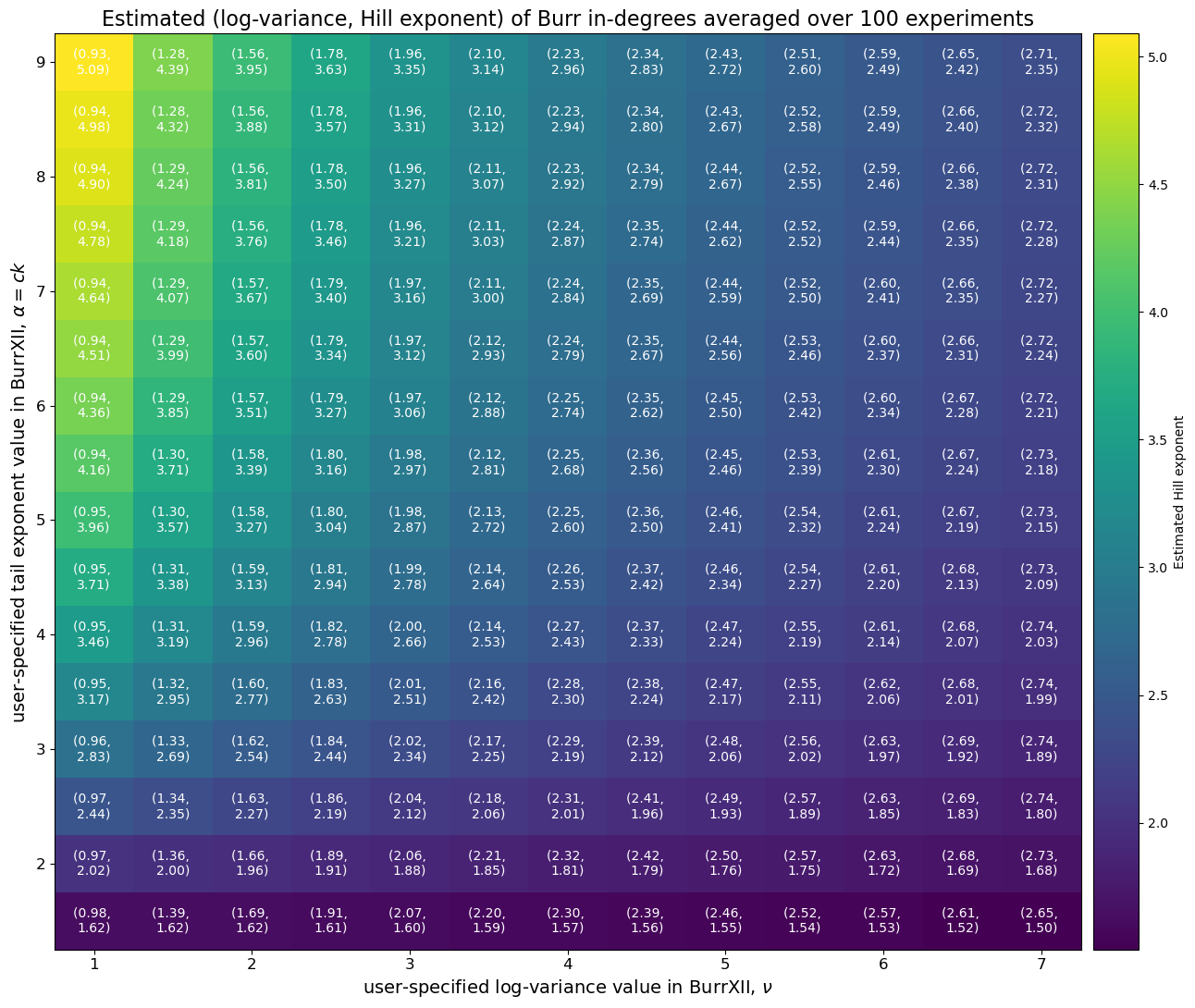}
    \caption{Meshgrid with step size $0.5$ over input tail exponent $\alpha$ and log-variance $\nu$ for in-degrees. For each grid point, we generate $5,000$ independent sequences of size $100,000$. The estimated Hill exponents and log-variances are displayed in white. We select $(\alpha, \nu) = (3,3)$ as input values, as they give an estimated Hill exponent of $2.34$ and log-variance of $2.02$. We deliberately select an $\alpha$ that yields a Hill exponent below the empirical value of $2.5$, since sampling adjacency matrices via the Configuration Model induces an upward bias, bringing the final Hill exponent in the synthetic network from $2.34$ to $2.5$. Since the Configuration Model does not affect the log-variance, $\nu$ is chosen to match the empirical value of $2.0$ as closely as possible.}
    \label{fig:burr_alpha_nu_in_degree_meshgrid}
\end{figure}

As for the in-degrees, we sample an in-degree sequence from a Burr XII distribution, then truncate and turn the observations to integers.
Specifically, we discard and resample any in-degree greater than $0.05N$. 
Similar to the out-degree case, Table \ref{tab:integerize_truncate_reconcile_degrees_hill001} shows that the realized Hill exponent and log-variance for in-degrees are lower than their user-defined values.
Accordingly, we calibrate the input parameters of the in-degree Burr XII distribution by trial and error. Figure \ref{fig:burr_alpha_nu_in_degree_meshgrid} presents a grid over input tail exponent $\alpha$ and log-variance $\nu$ for in-degrees. To generate an in-degree sequence in a network of $N=100,000$ firms with $\mu \approx 40$, $\alpha \approx 2.35$, $\nu \approx 2$, and a maximum in-degree capped at $0.05N$, we draw from a Burr XII distribution with parameters solved to match $\alpha = 3$ and $\nu=3$.

\subsection{Reconciling in- and out-degrees sums}
\label{subsubsection:reconcile_in_out_degree}
In a directed graph, the sum of in- and out-degrees are equal. In our procedure above, although our draws for the in- and out-degrees are designed to have similar means, they are highly unlikely to have exactly the same sum. We reconcile the totals by modifying the initial in-degree sequence to match the sum of out-degrees. Let
$$\Delta = \sum_{i}k^{\text{out}}_i - \sum_{j}k^{\text{in}}_j.$$
We draw $|\Delta| \geq 0$ independent indices from $\left\{1, \ldots, N\right\}$ with selection probabilities 
$$p_i = \frac{k_i^{\text{in}}}{ \sum_{j} k_j^{\text{in}}}, \quad i = 1 ,\ldots, N.$$
This is equivalent to sampling a multinomial random vector with $|\Delta|$ trials and cell probabilities $\{p_i\}_{i=1}^N$.
Let $(d_1,\ldots,d_N)$ denote the resulting nonnegative counts, with $\sum_{i=1}^{N} d_i = |\Delta|$.
If $\Delta>0$, we update $k^{\text{in}}_i \leftarrow k^{\text{in}}_i + d_i$ for each $i$; if $\Delta<0$, we update $k^{\text{in}}_i \leftarrow k^{\text{in}}_i - d_i$.
Equivalently, each of the $|\Delta|$ draws increments (when $\Delta>0$) or decrements (when $\Delta<0$) the selected firm's in-degree by one, so that the total in-degree changes by exactly $\Delta$.
Because decrements can concentrate on firms with small degrees when $\Delta<0$, some entries can become negative.
In such cases, or if any additional sampling constraints described below are violated, we discard the sequence and repeat the procedure by drawing a new truncated Burr in-degree sequence, and apply the reconciliation again. 

For networks with $N = 100,000$ firms, we only retain a reconciled in-degree sequence with tail exponent in the range $2.3 < \alpha < 2.4$, and\footnote{We use the same sampling constraints for $N=50,000$ firms as with $100,000$ firms. For $10,000$ firms, we use log-variance $\nu>1.85$.} log-variance $\nu > 1.95$. Unlike the out-degree sequence, we do not impose a lower bound for the maximum in-degree.

\subsection{Reordering Burr in-degrees and out-degrees}
\label{appendix_reorder_burr_in_out_degrees}
Since the firms' in- and out-degree sequences are drawn from independent Burr XII distributions, there is zero correlation between them.
We could order both distributions before matching them, so there would be a perfect correlation between in- and out-degrees. In practice, however, we want to match an empirical correlation of around $0.55$ \citep{bacilieri2026firm}. To do this, we match the ordered out-degree sequence not to the perfectly ordered in-degree sequence, but to an imperfectly ordered in-degree sequence. Specifically, we will obtain this imperfect ordering as the perfect ordering of a randomly scrambled sequence. In details, the procedure works as follows.

Given the in- and out-degree sequences, we rank every firm according to its in-degree,
$$
r_i = \text{rank}(k^{\text{in}}_i) \in \left\{0, \ldots,N-1 \right\}, \quad \tilde{r}_i = \frac{r_i}{N-1} \in [0,1].
$$
Then, we multiply each in-degree $k_i^{\text{in}}$ by a lognormal noise of unit mean and rank-dependent log-variance,
$$
k^{\text{in, scrambled}}_i = k^{\text{in}}_i \times \varepsilon_i, \text{ where } \varepsilon_i \sim \text{Lognormal}\left(-\frac{1}{2}(1-\tilde{r}_i)^{2\zeta}, (1-\tilde{r}_i)^{2\zeta}\right),
$$
noting that we use $\epsilon_i=1$ for the top ranked firm ($\tilde r_i=1$). Finally, we reorder the out-degree sequence to align with the \textit{scrambled} in-degrees' ranking. Therefore, the \emph{sorted} out-degrees are assigned to firms according to a permutation: the firm with smallest scrambled in-degree receives the smallest out-degree, the firm with the second-smallest scrambled in-degree receives the second-smallest out-degree, and so on. We calibrate the parameter $\zeta$ to ensure the Pearson correlation between the log in- and log out-degrees matches the empirical value of $0.55$ reported by \cite{bacilieri2026firm}. 
A reasonable value of $\zeta$ is $-0.14$. 

\subsection{Sampling the binary network}
With the two degree distributions in hand, we are ready to generate a binary network. Using the \texttt{igraph} package, we sample an adjacency matrix $\mathbf{A}$ from the configuration model with the in- and reordered out-degree sequences as inputs. 
We remove parallel edges and self-loops. 

This procedure induces an upward bias in the Hill exponents of in- and out-degrees computed from $\mathbf{A}$. To compensate for this, we deliberately choose Burr XII input parameters in Sections \ref{subsubsection:generate_burr_out_deg} and \ref{subsubsection:generate_burr_in_deg} that produce lower tail exponents prior to applying the configuration model ($1.4$ for out-degrees and $2.35$ for in-degrees). 
After sampling, the resulting tail exponents are approximately $1.5$ and $2.5$, closely matching the empirical targets. 

This concludes our binary network generation procedure. We now explain how we assign weights.

\section{Generating firm-level weighted networks}
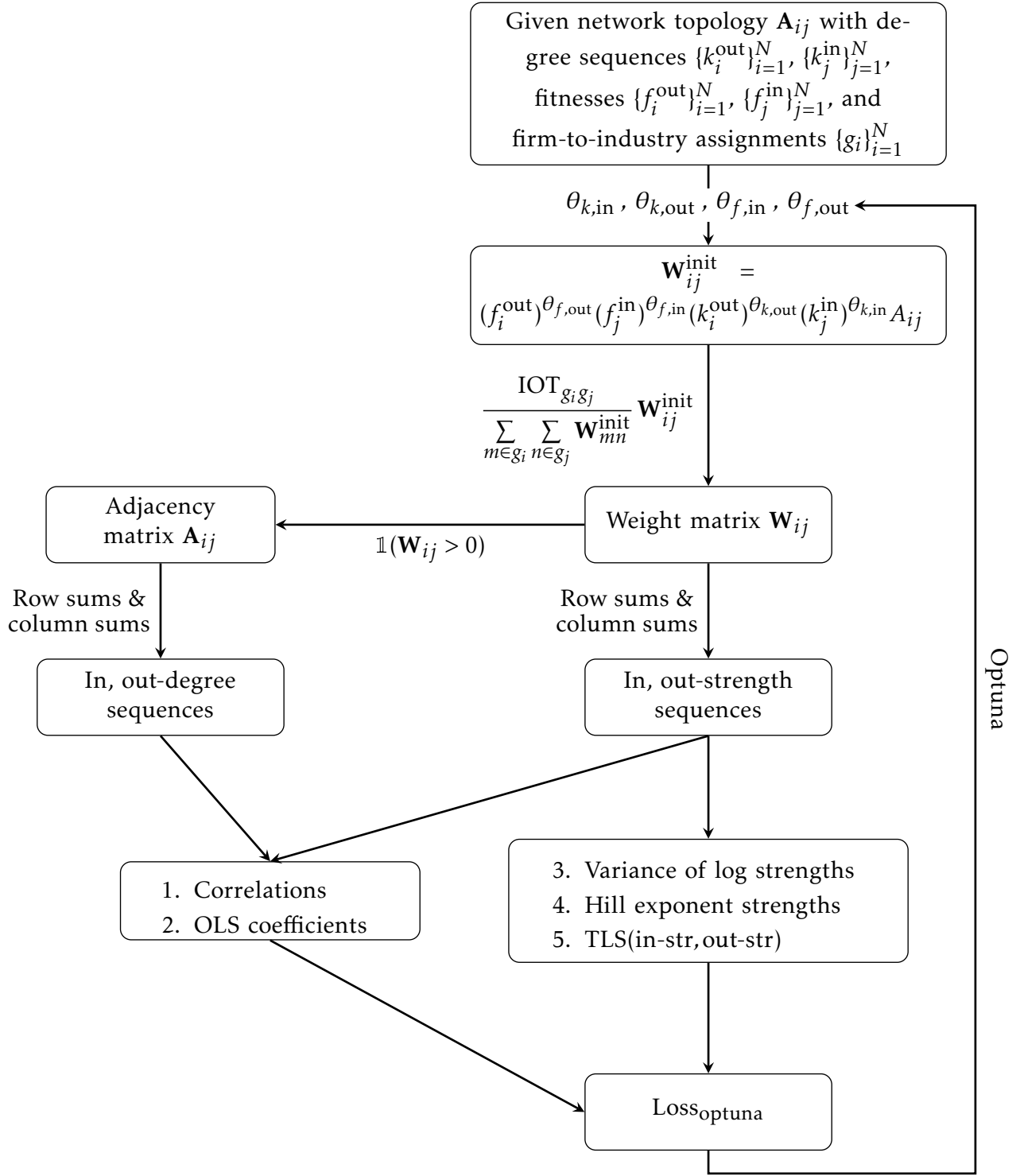
\begin{figure}[!htbp]
    \centering
    \resizebox{\linewidth}{!}{%
        \input{tables/model_architecture_tikz} 
    }
    \caption{Architecture for the optimization of parameters $\theta_{k,\mathrm{in}} \;,\;\theta_{k,\mathrm{out}}\;,\;\theta_{f,\mathrm{in}}\;,\;\theta_{f,\mathrm{out}}$ in the generation of synthetic supply networks. Here $\text{TLS}(y,x)$ represents the total least squares regression coefficient of $\ln y$ against $\ln x$.}
    \label{fig:pipeline_optuna}
\end{figure}

Conditional on the binary network generated, we generate weights for each existing edge. We do this by first generating ``fitnesses'' (latent variables for each node), and then using these and node degrees in a gravity-like formula to determine edge-level weights,
\begin{equation}
\mathbf{W}_{ij}^{\text{init}} = (f_{i}^{\text{out}})^{\theta_{f,\mathrm{out}}}(f_{j}^{\text{in}})^{\theta_{f,\mathrm{in}}} (k^{\text{out}}_i)^{\theta_{k,\mathrm{out}}} (k^{\text{in}}_j)^{\theta_{k,\mathrm{in}}}\mathbf{A}_{ij},
\label{eq:winit}
\end{equation}
where $f$ are the fitnesses and $\boldsymbol{\theta} = (\theta_{k,\mathrm{in}} \;,\;\theta_{k,\mathrm{out}}\;,\;\theta_{f,\mathrm{in}}\;,\;\theta_{f,\mathrm{out}})$ denotes a set of tunable parameters. We then rescale $\mathbf{W}_{ij}^{\text{init}}$ to match the input-output table (IOT). We first describe how we determine the fitnesses, followed by how we rescale to match the IOT. Figure \ref{fig:pipeline_optuna} outlines the model architecture.

\subsection{Generate node-level fitnesses}

Our goal is to introduce latent variables so that in Eq. \ref{eq:winit}, weights depend not only on degrees but also on an additional dimension of node heterogeneity. We have found that sampling the fitnesses from a log-normal distribution works well,
$$
\begin{pmatrix}
    \ln f^{\text{in}} \\ 
    \ln f^{\text{out}}
\end{pmatrix} 
\sim \mathcal{N}( \boldsymbol{\mu}_f, \boldsymbol{\Sigma}),
$$ 
where $\boldsymbol{\mu}_f$ and $\boldsymbol{\Sigma}$ denote the mean vector and covariance matrix of the log fitnesses. 

To determine its mean, note that because what matters is only heterogeneity, the mean of the latent variables will not impact the results substantially; we arbitrarily use $\boldsymbol{\mu}_f = (-8 \ln 10 , -8 \ln 10)$.

For the variance, note that the fitnesses enter Eq.~\ref{eq:winit} as $(f^{\mathrm{out}})^{\theta_{f,\mathrm{out}}}$ and $(f^{\mathrm{in}})^{\theta_{f,\mathrm{in}}}$. 
Thus, the variance of the powered fitness terms can be adjusted by changing $\theta_{f,\mathrm{out}}$ and $\theta_{f,\mathrm{in}}$, which are optimized later.
To facilitate parameter search, we want to ensure that all four $\theta$s are likely to be on similar scales. Intuitively, we expect the four $\theta$s to be on similar scales if the four variables are also on similar scales. Therefore, we assume that the variances of the log in- and log out-fitnesses are equal to the variances of the log in- and log out-degrees. Finally, for the covariance, we introduce a parameter $\rho$ that tunes the correlations. Summing up, we use
\begin{equation}
    \mathbf{\Sigma} = 
\begin{pmatrix}
    \mathrm{Var}( \ln k^{\mathrm{in}} ) && \rho*\sqrt{\mathrm{Var}( \ln k^{\mathrm{in}} )*\mathrm{Var}( \ln k^{\mathrm{out}} )} \\ 
    \rho*\sqrt{\mathrm{Var}( \ln k^{\mathrm{in}} )*\mathrm{Var}( \ln k^{\mathrm{out}} )} && \mathrm{Var}( \ln k^{\mathrm{out}} )
\end{pmatrix}.
\label{lognorm_str_cov}
\end{equation}
We set $\rho = 0.8$. After rescaling to match the IOT, the resulting log in- and log out-strength correlation is close to the empirical value of $0.5$.
Finally, to ensure that total in-fitness equals total out-fitness, we scale every out-fitness by the factor $\frac{\sum f^{\text{in}}}{\sum f^{\text{out}}}$.

Given the degrees and fitnesses, and assuming specific values for the parameter vector $\boldsymbol{\theta}$, we obtain the weighted adjacency matrix by direct calculation from Eq.~\ref{eq:winit}. Before we describe how we optimize for $\boldsymbol{\theta}$, we explain the final step of the process, assigning firms to industries and rescaling weights to ensure a match with IOTs.

\subsection{Rescaling a weighted network to match an input-output table}
\label{subsubsection:rescaling_factors}

We start from a firm-level weighted network $\mathbf{W}^{\text{init}}$ consisting of $N$ firms. We assume that the industries have a number of firms proportional to their industry sales. We first compute the preliminary industry count as
$$
\tilde{M}_s = \texttt{round} \left( N \frac{\sum\limits_{n=1}^{M} \text{IOT}_{sn}}{ \sum\limits_{m=1}^M \sum\limits_{n=1}^{M} \text{IOT}_{mn}} \right),
$$
where $\text{\text{IOT}}_{mn}$ denotes the monetary flow from industry $m$ to industry $n$, obtained from OECD input-output tables, and $\texttt{round}$ denotes rounding to the nearest integer.
Due to rounding, the sum $\sum_{s} \tilde{M}_s$ may be either smaller or larger than $N$.
We therefore obtain the adjusted industry counts $M_s$ by repeatedly adding one count to, or subtracting one count from, a uniformly randomly selected industry among those with positive current counts, until $\sum_s M_s = N$. We then assign firms sequentially according to these adjusted counts: the first $M_1$ firms are assigned to industry $1$, the next $M_2$ firms to industry $2$, and so on. This ensures that each firm is assigned to exactly one industry. Note that this way of assigning firms to industries is technically stochastic.

While proportional allocation appears as a reasonable assumption, our architecture is flexible enough to accommodate user-provided firm-to-industry mappings when such information is available.

With firms assigned to industries, we are ready to rescale the firm-to-firm weights to ensure that the industry-to-industry weights match a given IOT, using
$$
\mathbf{W}_{ij} = \frac{\text{IOT}_{{g_i}{g_j}}}{\sum\limits_{f \in g_i}\sum\limits_{h \in g_j} \mathbf{W}_{fh}^{\text{init}}} \mathbf{W}_{ij}^{\text{init}},
$$
where index $g_i$ corresponds to the industry of firm $i$. It is straightforward to check that the rescaled network $\mathbf{W}$ aggregates exactly into the IO table,  
$$
\sum\limits_{f \in g_i, h \in g_j} \mathbf{W}_{fh} = \text{IOT}_{g_i g_j}.
$$

For computational purposes, the rescaling can be written compactly in matrix notation. Defining the $M \times N$ aggregator matrix $\mathbf{S}$ as 
    \begin{equation*}
        \mathbf{S}_{uj} =
        \begin{cases} 
        1 & \text{if firm $j$ belongs to industry $u$} \\
        0 & \text{otherwise}
    \end{cases} \; ,
    \end{equation*} 
the rescaled weighted network is given by  
$$
\mathbf{W} = \mathbf{S}^{\top} \left( \text{IOT} \oslash \left[\mathbf{S}\mathbf{W}^{\text{init}} \mathbf{S}^{\top}\right] \right) \mathbf{S} \odot \mathbf{W}^{\text{init}},
$$
where $\oslash$ denotes Hadamard (element-wise) division, $\odot$ denotes the Hadamard (element-wise) product, and $\top$ denotes the transpose operation.
The industry-level aggregation of $\mathbf{W}^{\text{init}}$ is given by the matrix product $\mathbf{S}\mathbf{W}^{\text{init}} \mathbf{S}^{\top}$, see Chapter 4.9 of \cite{miller2009input}. 
The Hadamard division of $\text{IOT} \oslash \left[\mathbf{S}\mathbf{W}^{\text{init}} \mathbf{S}^{\top}\right]$ yields an $M \times M$ matrix of scaling factors, where each entry represents the ratio between the corresponding value in the empirical IOT and the respective value in the aggregated network $\mathbf{S}\mathbf{W}^{\text{init}} \mathbf{S}^{\top}$.
Finally, $\mathbf{S}^{\top} \left( \text{IOT} \oslash \left[\mathbf{S}\mathbf{W}^{\text{init}} \mathbf{S}^{\top}\right] \right) \mathbf{S}$ disaggregates the scaling factors into an $N \times N$ matrix, which are subsequently multiplied element-wise with the original weighted matrix $\mathbf{W}^{\text{init}}$: that is, we scale each element $(i,j)$ entry of $\mathbf{W}^{\text{init}}$ by the ratio between the $(g_i,g_j)$ entry of the specified IO table and the $(g_i, g_j)$ entry of the aggregation of $\mathbf{W}^{\text{init}}$.

This concludes the description of the full weight allocation pipeline, although we have so far assumed that the parameters were given. The next section describes one of the most important steps of our method, namely how we determine the values of $\theta$s such that the properties of weighted network will match empirical real-world properties.

\subsection{
Calibrating parameters \texorpdfstring{$\theta_{k,\mathrm{in}}, \theta_{k,\mathrm{out}},\theta_{f,\mathrm{in}}, \theta_{f,\mathrm{out}}$}
{theta parameters}}
\label{subsection:calibrating_parameters}

Calibrating the parameters $\boldsymbol{\theta} = (\theta_{k,\mathrm{in}} \;,\;\theta_{k,\mathrm{out}}\;,\;\theta_{f,\mathrm{in}}\;,\;\theta_{f,\mathrm{out}})$ allows the synthetic supply network to capture micro-level features while remaining exactly consistent with the macro-level input-output table. 
We calibrate these parameters using Optuna \citep{akiba2019optuna}, which is an open-source Python library for automated hyperparameter optimization that efficiently explores large parameter spaces by adaptively pruning unpromising candidates\footnote{
We have explored several alternatives, including manual grid search, which quickly became too expensive but was helpful for us to understand the landscape, and PyTorch (Adam), which in our case appeared more sensitive to initial conditions and delivered lower quality results, but has been helpful for us to understand trade-offs in the multi-objective loss function.%
}.
We minimize the loss function 
\begin{align}
\begin{split}
\text{Loss} &= \mathcal{D}(\text{hill}({\text{in-str}}))\\
& + \mathcal{D}(\text{hill(}{\text{out-str}})) \\
& + \mathcal{D}(\text{Corr}(\text{out-str}, \text{out-deg})) \\
& + \mathcal{D}(\text{Corr}(\text{in-str}, \text{in-deg})) \\
& + \mathcal{D}\text{(OLS(in-str, in-deg))} \\
& + \mathcal{D}\text{(OLS(in-deg,in-str))} \\
& + \mathcal{D}\text{(OLS(out-deg,out-str))} \\
& + \mathcal{D}\text{(OLS(out-str, out-deg))} \\
& + \mathcal{D}\text{(TLS(in-str, out-str))} \\
& + 10^{-3} \times \mathcal{D}(\text{Var}(\text{in-str})) \\
& + 10^{-3} \times \mathcal{D}(\text{Var}(\text{out-str}))
\end{split}
\label{eq:total_loss_definition}
\end{align}
where, for any statistic $q$, $\mathcal{D}(q) = \left[\widehat q(\boldsymbol{\theta})-q^\star\right]^2$ denotes the squared discrepancy between its value in the synthetic network, $\widehat q(\boldsymbol{\theta})$, and its empirical target, $q^{\star}$. Here, $\text{OLS}(y,x)$ and $\text{TLS}(y,x)$ denote the ordinary least squares and total least squares regressions of $y$ against $x$ respectively. Tail exponents are computed using the Hill estimator, $\text{Corr}$ denotes Pearson correlation, and OLS and TLS refer to ordinary and total least squares respectively. All correlations, OLS and TLS coefficients, and variance terms are computed on log-transformed variables. All loss components are equally weighted, except for the strengths variances, which receive a weight of $10^{-3}$.
Figure \ref{fig:pipeline_optuna} illustrates this optimization procedure.

To be clear, the $\theta$ parameters affect only the weights assigned to existing edges, while the network topology $\mathbf{A}_{ij}$ remains fixed. Thus, this optimization does not include terms targeting binary network properties.

We run Optuna for a fixed number of trials, which determines the search duration. In our implementation, we use $200$ trials and search over the parameter space $$\theta_{k,\mathrm{in}} \;,\;\theta_{k,\mathrm{out}}\;,\;\theta_{f,\mathrm{in}}\;,\;\theta_{f,\mathrm{out}} \in [-2,2]^4,$$ which we find is sufficiently broad to identify optimal parameters\footnote{As a robustness check, we searched over different parameter spaces in Optuna, the widest being $[-4,4]^4$, and obtained the same behavior of optimal parameters.}. On a machine with an \texttt{Intel Xeon Gold} $6226$R processor ($12$ cores, $2.90$ GHz) and $125$GB RAM, a single search procedure takes 7 minutes.

\begin{figure}[!htbp]
    \centering
    \includegraphics[width=\linewidth]{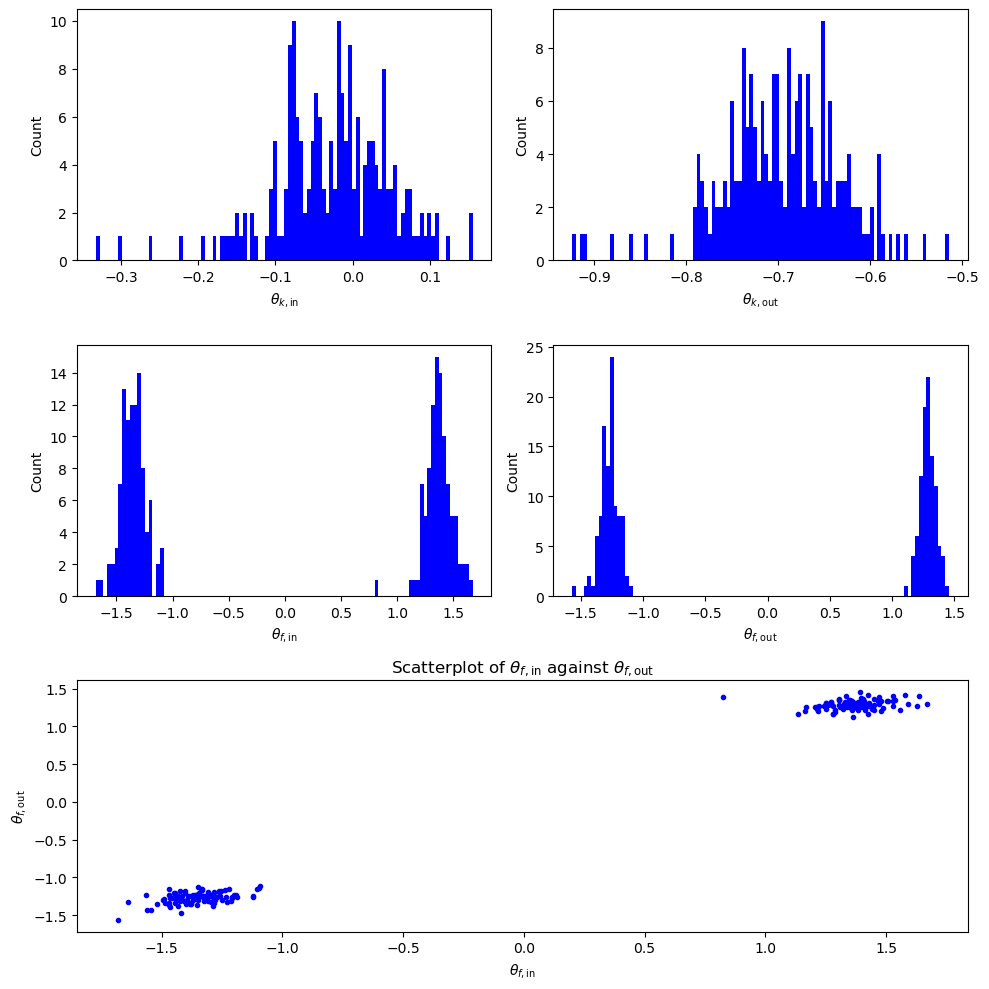}
    \caption{Distribution of optimal parameters identified using Optuna. The parameter search procedure is repeated $200$ times, and the histograms report the optimal values obtained in each run. The estimates of $\theta_{k,\mathrm{in}}$ are centered around zero, while $\theta_{k,\mathrm{out}}$ is consistently negative. $\theta_{f,\mathrm{in}}$ and $\theta_{f,\mathrm{out}}$ exhibit bimodal distributions. The scatterplot in the bottom panel shows two distinct clusters, indicating that $\theta_{f,\mathrm{in}}$ and $\theta_{f,\mathrm{out}}$ are either both negative or both positive: positive optimal parameters occurred $101$ times and negative parameters $99$, i.e. both cases occur equally often. The means of these optimal parameters are reported in Table \ref{tab:optima_parameters_from_optuna_stats}.}
    \label{fig:optimal_parameters_from_optuna}
\end{figure}

Figure \ref{fig:optimal_parameters_from_optuna} illustrates the best parameter tuples by repeating the parameter search procedure $200$ times, where each search consists of $200$ trials. 
Table \ref{tab:optima_parameters_from_optuna_stats} presents the mean and standard deviation of the best parameters. 

We observe several peculiarities: the optimal value of $\theta_{k,\mathrm{in}}$ is centered around $0$, while $\theta_{k,\mathrm{out}}$ is consistently negative, with a relatively precise estimate around the interval $[-0.8, -0.6]$. Instead, the parameters appear relatively precisely estimated, baring an identification problem: $\theta_{f,\mathrm{in}}$ and $\theta_{f,\mathrm{out}}$ are either both negative or both positive, with equal frequency, and similar means. 

\input{tables/mean_std_of_best_theta_from_optuna_search}

Given these results, in the main paper, we use the values
$(\theta_{k,\mathrm{in}} \;,\;\theta_{k,\mathrm{out}}\;,\;\theta_{f,\mathrm{in}}\;,\;\theta_{f,\mathrm{out}}) = (0, -0.7, -1.3, -1.3)$. 

Note that these results are likely to be of independent interest, as other researchers could use our param\-etrization of the latent variables, Eq.~\ref{eq:winit}, and these values of $\theta$s to assign weights to binary networks obtained using different methods, and likely achieve a good match to desirable weighted network properties. 

To investigate more carefully whether these values (and our entire pipeline) work well in other context, the next section shows the robustness of the result to networks of different sizes, and calibrated on different IOTs.

\section{Robustness checks}
We assess the robustness of our methodology in the following ways:
\begin{enumerate}
    \item \textbf{Robustness against stochastic variation in the pipeline}: Generate synthetic networks using the optimal parameters identified by Optuna, where each network uses a different realization of degree sequences, fitnesses, adjacency matrices $\mathbf{A}_{ij}$, and firm-to-industry assignments.
    \item \textbf{Robustness against small changes in $\boldsymbol{\theta}$ parameter values}: Repeat the above by using the optimal parameters rounded to one decimal place.
    \item \textbf{Robustness against choice of country.} Using the rounded optimal parameters, generate synthetic networks for different country input-output tables.
    \item \textbf{Robustness against choice of number of firms.} Using the rounded optimal parameters, generate synthetic networks of varying network sizes.
\end{enumerate}

\subsection{Robustness of generated networks}
\label{appendix_robustness_of_gen_networks}
Table \ref{tab:results_stability} reports the network properties for the first two robustness checks, based on $100$ networks of size $100,000$ per parameter setting using the $2015$ Hungarian input-output table. The column ``Positive'' corresponds to networks generated with positive fitness parameters, $$( \theta_{k,\mathrm{in}} \;,\;\theta_{k,\mathrm{out}}\;,\;\theta_{f,\mathrm{in}}\;,\;\theta_{f,\mathrm{out}} ) =(-0.0265,-0.697,1.37,1.29),$$ while ``Negative'' corresponds to networks generated using negative fitness parameters, $$( \theta_{k,\mathrm{in}} \;,\;\theta_{k,\mathrm{out}}\;,\;\theta_{f,\mathrm{in}}\;,\;\theta_{f,\mathrm{out}}) = (-0.0265,-0.697,-1.35,-1.27).$$ 
The column ``Rounded negative'' uses the rounded negative parameters,
$$( \theta_{k,\mathrm{in}} \;,\;\theta_{k,\mathrm{out}}\;,\;\theta_{f,\mathrm{in}}\;,\;\theta_{f,\mathrm{out}}) = (0, -0.7, -1.3, -1.3).$$

\begin{figure}[!htbp]
    \centering
    \includegraphics[width=0.75\columnwidth, keepaspectratio]{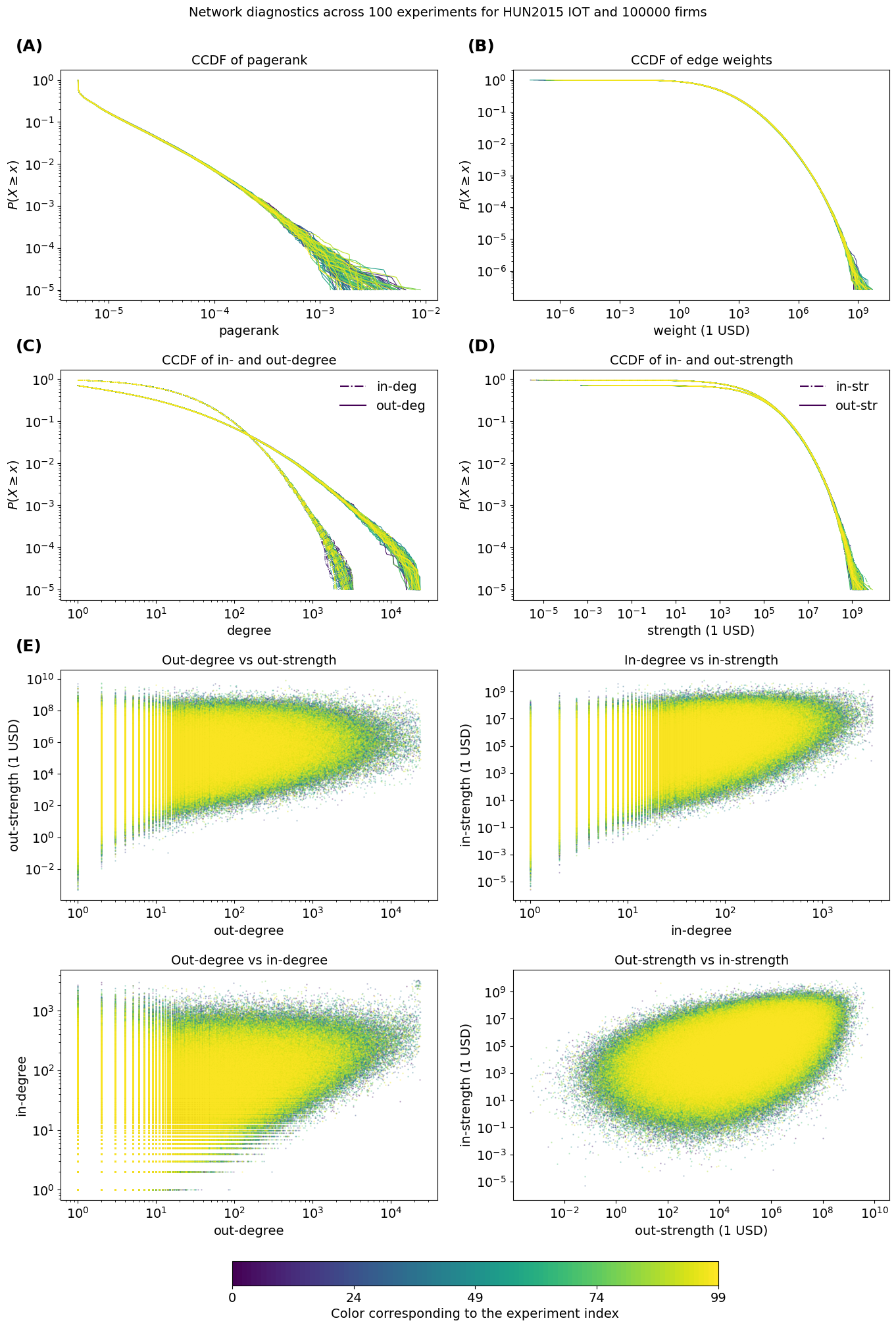}
    \caption{Distributional properties of $100$ networks generated using $(\theta_{k,\mathrm{in}} \;,\;\theta_{k,\mathrm{out}}\;,\;\theta_{f,\mathrm{in}}\;,\;\theta_{f,\mathrm{out}}) = (-0.0265,-0.697,1.37,1.29)$, with $\theta_{f,\mathrm{in}}\;,\;\theta_{f,\mathrm{out}}$ both positive. Randomness is introduced from different degree sequences, fitnesses, draw of adjacency matrix from the configuration model, and firm-to-industry assignment. The Figure shows that the networks are very similar across experiments. (A) Complementary cumulative distribution function (CCDF) of PageRank. (B) CCDF of edge weights. (C) CCDF of in- and out-degrees. (D) CCDF of in- and out-strengths. (E) 2D histograms of the four joint distributions consisting of out-strength against out-degree, in-strength against in-degree, in-degree against out-degree, and out-strength against in-strength.}
    \label{fig:diagnostics_wgt_mat_init_positive_thetas_hun100k_100networks}
\end{figure}

Figures \ref{fig:diagnostics_wgt_mat_init_positive_thetas_hun100k_100networks} shows the corresponding distributional properties for the ``positive'' case (very similar results hold for the other two). 

\input{tables/100_networks_using_pos_neg_roundedneg_thetas}

Across runs, network properties have low variation, with standard deviations typically one or even two orders of magnitude smaller than their means. For example, under the column ``Negative'', the in-degree Hill exponent has a mean of $2.54$ and a standard deviation of $0.03$, close to the target value of $2.5$.

Moreover, consistent with empirical evidence from Value-Added Tax (VAT) data \citep{bacilieri2026firm}, the asymmetry between the in-degree and out-degree complementary cumulative distribution functions (CCDFs) is clearly evident, with a maximum out-degree approximately an order of magnitude larger than the maximum in-degree.

In contrast, the in-strength and out-strength CCDFs are indistinguishable, again consistent with \cite{bacilieri2026firm}. 

The panel of $2$D histograms further confirm the stability of joint distributions across experiments.

Overall, the results demonstrate that, for a fixed set of parameters $(\theta_{k,\mathrm{in}} \;,\;\theta_{k,\mathrm{out}}\;,\;\theta_{f,\mathrm{in}}\;,\;\theta_{f,\mathrm{out}})$, the resulting distributional and network properties remain stable across different realizations of degree sequences, fitnesses, draws from the Configuration Model, and firm-to-industry assignments. 
This robustness implies that the calibrated parameters can be reused to generate synthetic networks without running Optuna. 
For ease of use, we provide default values of $(\theta_{k,\mathrm{in}} \;,\;\theta_{k,\mathrm{out}}\;,\;\theta_{f,\mathrm{in}}\;,\;\theta_{f,\mathrm{out}}) = (0,-0.7,-1.3,-1.3)$ in the codebase.

\subsection{Robustness across different countries' IOT}
We examine the stability of networks generated for different country input-output tables, namely Belgium, Hungary, and Brunei Darussalam for the year $2015$. 
For each country, we generate $40$ networks of $100,000$ firms using the rounded negative parameters $(\theta_{k,\mathrm{in}} \;,\;\theta_{k,\mathrm{out}}\;,\;\theta_{f,\mathrm{in}}\;,\;\theta_{f,\mathrm{out}}) = (0,-0.7,-1.3,-1.3)$.

We choose Belgium and Hungary because their firm-level VAT dataset has been well-studied \cite{bacilieri2026firm}. We chose Brunei Darussalam to provide maximum contrast: its IOT exhibits the largest deviation from Hungary among the available OECD datasets\footnote{
Measured by the column-normalized $\ell_1$ distance $\sum_{m=1}^{45}\sum_{n=1}^{45} \left| \frac{\text{IOT}_{mn}}{\sum_{m=1}^{45}\text{IOT}_{mn}} - \frac{\text{IOT}_{mn}^{\text{Hungary}}}{\sum_{m=1}^{45} \text{IOT}^{\text{Hungary}}_{mn}} \right|$.
}.

\begin{figure}[!htbp]
    \centering
    \includegraphics[width = 0.87\columnwidth, keepaspectratio]{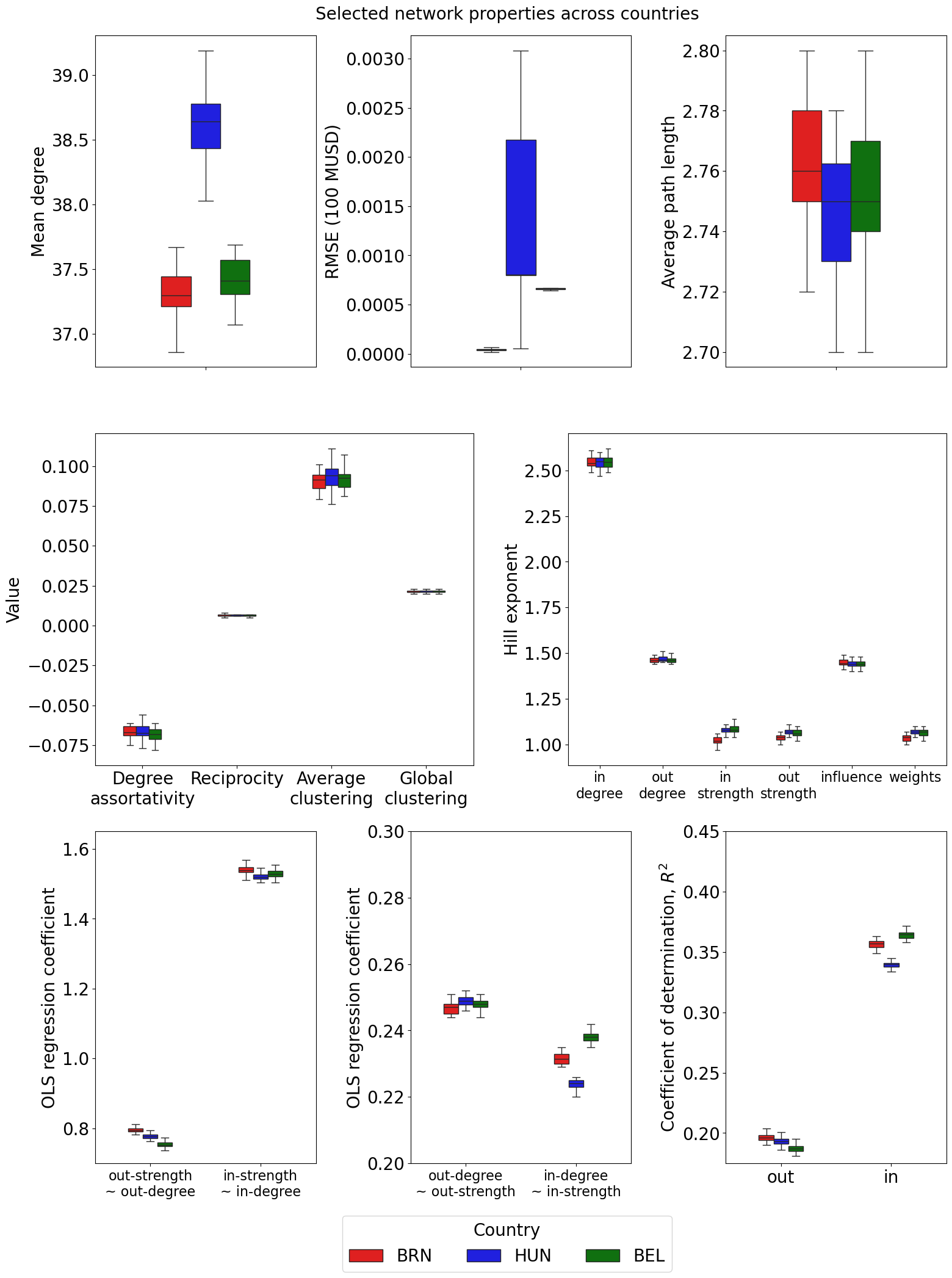}
    \caption{Network properties of generated networks that match the $2015$ input-output tables of Brunei Darussalam, Hungary, and Belgium. For each country, $40$ networks with $100,000$ firms are generated using $(\theta_{k,\mathrm{in}} \;,\;\theta_{k,\mathrm{out}}\;,\;\theta_{f,\mathrm{in}}\;,\;\theta_{f,\mathrm{out}}) = (0,-0.7,-1.3,-1.3)$, and the resulting properties are shown using boxplots. Network properties exhibit minimal variation across countries. Cross-country differences are visible but remain small relative to the relevant targets and scales: the mean degree stays close to the target value of $40$, and the aggregated matrix has low RMSE relative to the scale of the input-output table.}
    \label{fig:stability_across_countries}
\end{figure}

As illustrated in Figure \ref{fig:stability_across_countries}, we find that the network properties are similar between countries. 
This demonstrates that our methodology generalizes across countries while preserving key empirical firm-level characteristics.
By construction, all synthetic networks aggregate closely to their target input-output tables. 

\subsection{Robustness across different network sizes}
\begin{figure}[!htbp]
    \centering
    \includegraphics[height=0.83\textheight, keepaspectratio]{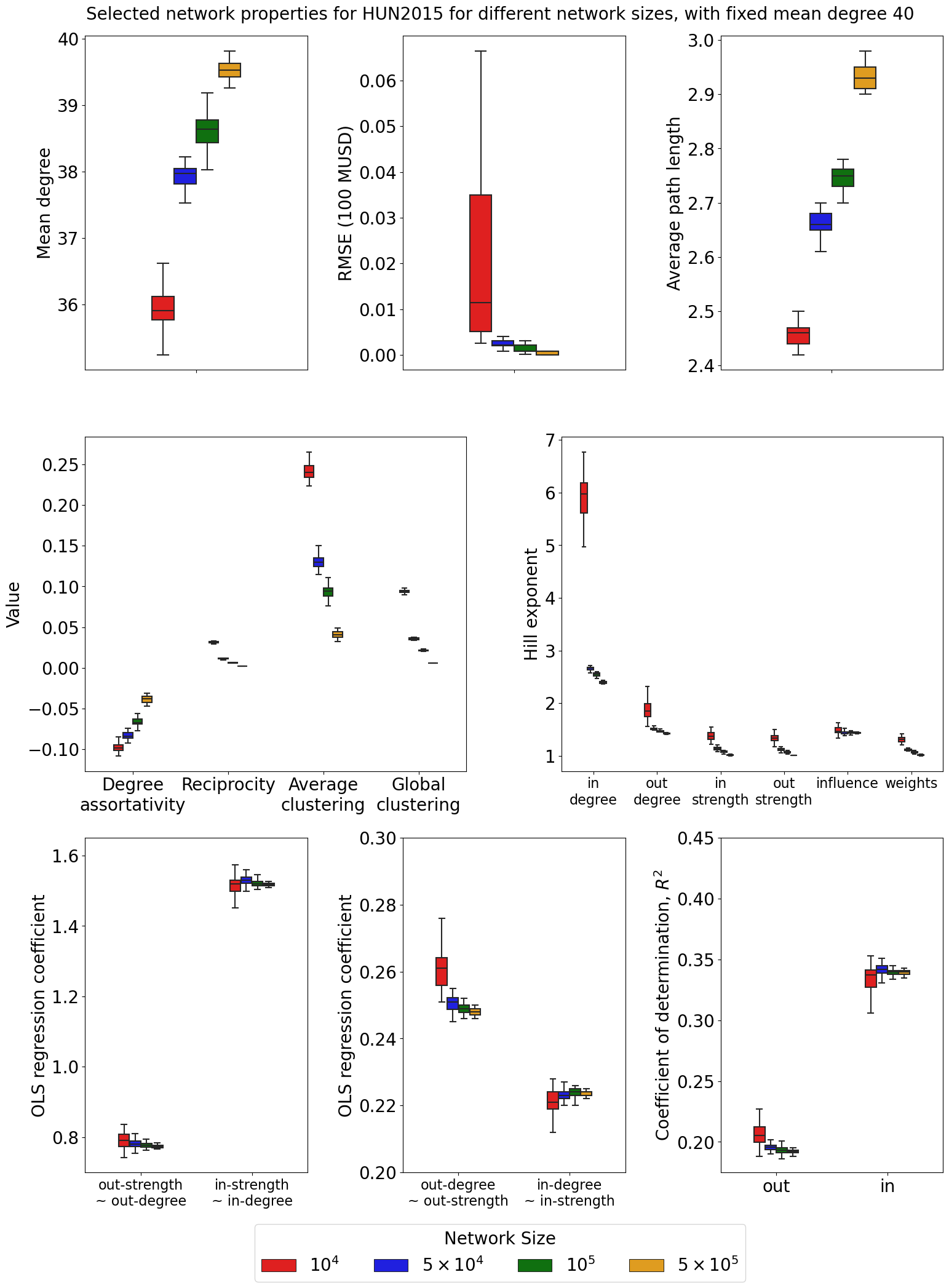}
    \caption{Network properties across different network sizes. Synthetic networks match the $2015$ Hungarian input-output table. For each network size, we fix the mean degree parameter $\mu$ at $40$ and generate $40$ independent networks using the rounded negative parameters $(\theta_{k,\mathrm{in}} \;,\;\theta_{k,\mathrm{out}}\;,\;\theta_{f,\mathrm{in}}\;,\;\theta_{f,\mathrm{out}}) = (0,-0.7,-1.3,-1.3)$. Properties are summarized using boxplots. Systematic variation arises from discretization, truncation, and reconciliation of Burr XII draws. Smaller networks have upward-biased degree Hill exponents and greater deviation from the target mean degree. As network size increases, average path length rises, while degree disassortativity (in absolute value), reciprocity, clustering, and Hill exponents of degrees, strengths, influence, and weights decline. OLS coefficients remain stable across network sizes.}
    \label{fig:stability_across_network_size}
\end{figure}

Figure \ref{fig:stability_across_network_size} examines how network properties vary with network size for $10,000$, $50,000$, $100,000$, and $500,000$ firms. 
For each network size, we generate $40$ networks using the rounded negative parameters $(\theta_{k,\mathrm{in}} \;,\;\theta_{k,\mathrm{out}}\allowbreak \;,\;\theta_{f,\mathrm{in}}\;,\;\theta_{f,\mathrm{out}}) = (0,-0.7,-1.3,-1.3)$, where all networks match the $2015$ Hungarian input-output table. 
We fix the mean degree parameter at $40$, that is, we set $\mu=40$ when sampling Burr degrees (see Eq. \eqref{eq:c_and_lambda_formula}).

As network size increases, degree disassortativity (in absolute value), reciprocity, average clustering, and global clustering all decline. 
This is expected because adjacency matrices are sampled from the Configuration Model, under which the global clustering coefficient scales inversely with network size \citep{newman2018networks}.

Two systematic patterns emerge in smaller networks: upward-biased degree Hill exponents and greater deviation of the mean degree from its target.
Both are explained by the ablation study in Table \ref{tab:integerize_truncate_reconcile_degrees_hill001}. 
For a fixed mean of $40$, the block ``Burr with discretization, truncation, and reconciliation'' shows that, as network size decreases from $100,000$ to $1,000$, the Hill exponent increases from $1.4$ to $5.83$ for out-degrees and from $2.33$ to $24.7$ for in-degrees.
This reflects the use of a fixed tail probability $p=0.01$ when estimating the Hill exponents of degree distributions, which leads to fewer tail observations in smaller networks and hence an upward bias. 
For the same range of network sizes and fixed mean degree parameter $\mu=40$, the mean degree falls from $39$ to $19.6$, indicating increasing deviation. 

In contrast, regression-based properties remain stable across network sizes, indicating that the optimized parameters $(\theta_{k,\mathrm{in}} \;,\;\theta_{k,\mathrm{out}}\;,\;\theta_{f,\mathrm{in}}\;,\;\theta_{f,\mathrm{out}})$ generalize well across scales. 

All synthetic networks aggregate closely to the $2015$ Hungarian IOT by construction, although aggregation accuracy improves as the network size increases. 

Taken together, these results highlight the importance of choosing an appropriate network size to reproduce the key empirical firm-level characteristics, and motivate representing economic models at a $1:1$ scale.

\newpage
\section{Proofs}
\begin{lemma}
\label{lemma_uniform_beta}
    Let $U$ be a Uniform$[0,1]$ random variable. For $k > 0$, we have $1 - (1-U)^{\frac{1}{k}} \overset{d}{=} \text{Beta}(1,k)$.
    \begin{proof}
        Recall that the density function of a Beta$(1,k)$ random variable with support $x\in[0,1]$ is given by $k(1-x)^{k-1}$.
        Then we have, 
        \begin{align*}
            \mathbb{P}\left( 1 - (1 -U)^{\frac{1}{k}} \leq u\right) &= \mathbb{P}\left(U \leq 1 - (1-u)^k \right)
            \\ 
            &= 1 - (1-u)^k.
        \end{align*}
        Differentiating with respect to $u$, the density function of $1 - (1-U)^{\frac{1}{k}}$ is $k(1-u)^{k-1}$.
    \end{proof}
\end{lemma}

\begin{proposition}
    Let $X$ be a BurrXII$(c,k,\lambda)$ random variable with strictly positive parameters $(c,k,\lambda)$, $V$ be a Beta$(1,k)$ random variable, and $U$ be a Uniform$[0,1]$ random variable. Then $\left(\frac{X}{\lambda}\right)^c \overset{d}{=} \frac{V}{1-V}$.
    \begin{proof}
        By Eq. \eqref{eq:burr_from_uniform_rv} and Lemma \ref{lemma_uniform_beta}, we have 
        \begin{align*}
            \frac{V}{1-V} = \frac{1 - (1-U)^{\frac{1}{k}}}{1 - \left[ 1 - (1-U)^{\frac{1}{k}} \right]} = \frac{1 - (1-U)^{\frac{1}{k}} }{(1-U)^{\frac{1}{k}}} = (1-U)^{-\frac{1}{k}} -1 = \left(\frac{X}{\lambda} \right)^{c}.
        \end{align*}
    \end{proof}
\end{proposition}

\begin{corollary}
\label{corollary:soln_for_mean_and_logvar}
Denote a Burr XII random variable with strictly positive parameters $(c,k,\lambda)$ by $X \sim \text{BurrXII}(c,k,\lambda)$ and suppose that $ck > 1$, then its expectation is 
$$
\mathbb{E} \left[X \right] = \lambda k B \left(k - \frac{1}{c}, 1 + \frac{1}{c}\right),
$$
for $B(\cdot,\cdot)$ is the Beta function. Furthermore, the log-variance is given by $$\text{Var}\left(\ln X\right) = \frac{\frac{\pi^2}{6} + \psi_1(k)}{c^2},$$ where $\psi_1(\cdot)$ is trigamma function. 
\begin{proof}
    For $V \sim \text{Beta}(1,k)$, we have 
    $$ \left(\frac{X}{\lambda}\right)^c \overset{d}{=} \frac{V}{1-V} \iff X \overset{d}{=} \lambda \left( \frac{V}{1-V}\right)^{\frac{1}{c}},$$
    then taking expectations, we have
    \begin{align*}
        \mathbb{E}\left[X\right] &= \lambda \mathbb{E} \left( \frac{V}{1-V}\right)^{\frac{1}{c}} 
        \\
        &= \lambda \int_{0}^1 \left(\frac{v}{1-v}\right)^{\frac{1}{c}} k \left(1-v\right)^{k-1} dv 
        \\
        &= \lambda k \int_0^1 v^{\frac{1}{c}} \left( 1 - v \right)^{k - 1 - \frac{1}{c}} dv
        \\
        &= \lambda k B\left( k - \frac{1}{c} , 1 + \frac{1}{c} \right).
    \end{align*}
    For the log-variance, taking log on both sides of $\left(\frac{X}{\lambda}\right)^c \overset{d}{=} \frac{V}{1-V}$, we have 
    \begin{align*}
        \ln X &= \ln \lambda + \frac{1}{c}\left( \ln V - \ln \left(1-V\right)\right)
        \\
        \implies \text{Var}\left(\ln X\right) &= \frac{\text{Var}\left(\ln V - \ln \left(1-V\right) \right) }{c^2} 
        \\
        &= \frac{\text{Var} \ln V + \text{Var} \ln\left(1-V\right) - 2\text{Cov}\left(\ln V, \ln\left(1-V\right) \right)}{c^2}
        \\
        &= \frac{ \frac{\pi^2}{6} + \psi_1(k)}{c^2},
    \end{align*} 
where we used the properties $\psi_1(1) = \frac{\pi^2}{6}$, Var $\ln V = \psi_1(1) - \psi_1(1+k)$,  Var $\ln \left(1-V\right) = \psi_1(k) - \psi_1(1+k)$, and $\text{Cov}\left( \ln V , \ln \left(1 - V\right) \right) = - \psi_1(1+k)$ for the final equality. 
\end{proof}
\end{corollary}

\begin{corollary}
\label{corollary:log-log_slope_of_burr}
Denote a Burr XII random variable with strictly positive parameters $(c,k,\lambda)$ by $X \sim \text{BurrXII}(c,k,\lambda)$,  whose Cumulative Distribution Function (CDF) is given by $F(x) = 1 - \left[1+\left(\frac{x}{\lambda}\right)^c\right]^{-k}$. Define $\alpha = ck$ and assume $\alpha > 1$. Denote its Complementary Cumulative Distribution Function (CCDF) by $\bar{F}(x) = 1 - F(x) = \mathbb{P}\left(X \geq x\right)$. Define the point $x_p$ such that $\bar{F}(x_p) = p$. Then the $\log$-$\log$ slope of its Complementary Cumulative Distribution Function (CCDF) at point $x_p$ is given by 
$$\alpha\left(1 - p^{\frac{1}{k}}\right).$$
\begin{proof}
    Denote the $\log$-$\log$ slope of the CCDF by $$\beta(x):=-\frac{d\log \bar F(x)}{d\log x}.$$
    Since $\log \bar F(x)=-k\log\left(1+\left(\frac{x}{\lambda}\right)^c\right)$, we have 
    $$
    \frac{d\log \bar F(x)}{d\log x} =- \alpha\,
    \frac{\left(\frac{x}{\lambda}\right)^c}{1+\left(\frac{x}{\lambda}\right)^c},
    $$
    so \begin{equation}
    \beta(x)= \alpha\frac{\left(\frac{x}{\lambda}\right)^c}{1+\left(\frac{x}{\lambda}\right)^c}.
    \label{eq:local-slope}
    \end{equation}
    Observe that Eq. \eqref{eq:local-slope} has asymptotic slope $\alpha$ as $x\to \infty$.
    Because $\bar{F}(x_p) = p$, we have 
    $$\left(\frac{x_p}{\lambda}\right)^c = p^{-1/k}-1.$$
    Substituting into Eq. \eqref{eq:local-slope} at $x = x_p$:
    $$
    \beta(x_p) = \alpha \left(1 - p^{\frac{1}{k}}\right).
    $$
    Lastly, from Corollary \ref{corollary:soln_for_mean_and_logvar}, this equation is also equivalent to 
    $$\beta(x_p) = k \sqrt{\frac{\frac{\pi^2}{6} + \psi_1(k)}{\nu}} \left(1 - p^{\frac{1}{k}}\right),$$
    which shows that the $\log$-$\log$ slope depends on the log-variance $\nu$.
\end{proof} 
\end{corollary}

%% file: tables/generation_procedure_in_box.tex
\begin{blackbox}[Initialization procedure]
\begin{enumerate}[leftmargin=*, itemsep=0.25em]
    \item Generate a sequence of (continuous) out-degrees sampled from a Burr XII distribution. Discard and resample observations that are too large. Convert these observations to integers.  
    \item Generate a sequence of in-degrees sampled from another Burr XII distribution. Discard and resample observations that are too large. Convert the observations to integers. Reconcile the totals so that the sum of the in-degrees equals the sum of out-degrees.  
    \item Reorder the out-degrees to create a correlation with the in-degree sequence. 
    \item Sample an adjacency matrix $\mathbf{A}_{ij}$ from the Configuration Model based on the degree sequences $\{k^{\text{in}}\}_{i=1}^{N}$ and $\{k^{\text{out}}\}_{i=1}^{N}$ defined above. 
    \item Sample jointly lognormal in- and out-fitnesses $\{f^{\text{in}}\}_{i=1}^{N}$ and $\{f^{\text{out}}\}_{i=1}^{N}$ with specific mean and covariance. Multiply out-fitnesses by $\frac{\sum f^{\text{in}}}{\sum f^{\text{out}}}$ to ensure $\sum f^{\text{out}} = \sum f^{\text{in}}$.
    \item Calibrate the parameters $\theta_{k,\mathrm{in}} \;,\;\theta_{k,\mathrm{out}}\;,\;\theta_{f,\mathrm{in}}\;,\;\theta_{f,\mathrm{out}}$ using Optuna by minimizing a loss function (Eq.~ \ref{eq:total_loss_definition} defined below). Denote the optimized parameters as $\theta_{k,\mathrm{in}}^{*} \;,\; \theta_{k,\mathrm{out}}^{*} \;,\; \theta_{f,\mathrm{in}}^{*} \;,\; \theta_{f,\mathrm{out}}^{*}$.
    \item Compute an initial set of weights using
    \begin{align*}
    \mathbf{W}_{ij}^{\text{init},*} &= (f_{i}^{\text{out}})^{\theta^{*}_{f,\mathrm{out}}}(f_{j}^{\text{in}})^{\theta^{*}_{f,\mathrm{in}}} (k^{\text{out}}_i)^{\theta^{*}_{k,\mathrm{out}}} (k^{\text{in}}_j)^{\theta^{*}_{k,\mathrm{in}}}\mathbf{A}_{ij}.
    \end{align*}
    \item Obtain the final weights by rescaling the initial weights so that they agree with the aggregate Input-Output Tables, using
    \begin{align*}
    \mathbf{W}^{*}_{ij} &= \frac{\text{IOT}_{{g_i}{g_j}}}{\sum\limits_{f \in g_i}\sum\limits_{h \in g_j} \mathbf{W}_{fh}^{\text{init},*}} \mathbf{W}_{ij}^{\text{init},*}.
        \end{align*}

\end{enumerate}
\end{blackbox}

%% file: tables/choosing_target_val.tex
\begin{table}[!htbp]
\centering
\small
\begin{threeparttable}
\begin{tabular}{@{}lrrlrr@{}}
\toprule
Property & Target value & Source & Property & Target value & Source \\ 
\cmidrule(lr){1-3} \cmidrule(lr){4-6}

Mean degree & $0.86N^{1/3}$ & S1 & Tail exponent $k^{\mathrm{in}}$ & 2.5 & S5 \\
Mean of log $k^{\mathrm{in}}$ & 3 & S2 & Tail exponent $k^{\mathrm{out}}$ & 1.5 & S5 \\
Mean of log $k^{\mathrm{out}}$ & 2 & S2 & Tail exponent $s^{\mathrm{in}}$ & 1 & S4 \\
Variance of log $k^{\mathrm{in}}$ & 2 & S2 & Tail exponent $s^{\mathrm{out}}$ & 1 & S4 \\
Variance of log $k^{\mathrm{out}}$ & 3 & S2 & Tail exponent weights & 1.15 & S5 \\
Maximum $k^{\mathrm{in}}$ & At most 5\% of $N$ & -- & Tail exponent influence & $[1.2,1.3]$ & S4 \\
Maximum $k^{\mathrm{out}}$ & At most 40\% of $N$ & -- & Tail prob. Hill degrees & 0.01 & S6 \\
Mean of log $s^{\mathrm{in}}$ & 10 & S3 & Tail prob. Hill strengths & 0.02 & S7 \\
Mean of log $s^{\mathrm{out}}$ & 10 & S3 & Tail prob. Hill weights & 0.0015 & S8 \\
Variance of log $s^{\mathrm{in}}$ & 9 & S3 & Tail prob. Hill influence & 0.04 & S9 \\
Variance of log $s^{\mathrm{out}}$ & 8 & S2 & Corr: $k^{\text{in}}\sim k^{\text{out}}$ & 0.55 & S10 \\
Average path length & 3 & S4 & Corr: $s^{\text{in}}\sim s^{\text{out}}$ & 0.5 & S10 \\
Degree assortativity & $[-0.2,-0.01]$ & S4 & Corr: $s^{\text{out}}\sim k^{\text{out}}$ & 0.5 & S11 \\
Reciprocity & $[0.03,0.05]$ & S4 & Corr: $s^{\text{in}}\sim k^{\text{in}}$ & 0.75 & S11 \\
Average clustering & $[0.19,0.28]$ & S4 & OLS: $s^{\text{out}}\sim k^{\text{out}}$ & 0.76 & S11 \\
Global clustering & low & S4 & OLS: $k^{\text{out}}\sim s^{\text{out}}$ & 0.33 & S12 \\
& & & OLS: $s^{\text{in}}\sim k^{\text{in}}$ & 1.4 & S11 \\
& & & OLS: $k^{\text{in}}\sim s^{\text{in}}$ & 0.4 & S12 \\
& & & TLS: $s^{\text{in}}\sim s^{\text{out}}$ & 1 & S4 \\
& & & TLS: $k^{\text{in}}\sim k^{\text{out}}$ & 0.7 & S5 \\

\bottomrule
\end{tabular}

\begin{tablenotes}
\footnotesize
\item[S1] From \cite{bacilieri2026firm} and \cite{pichler2023building}.
\item[S2] Rounded to nearest integer from Table B.10 of \cite{bacilieri2026firm}; Ecuador (2015) and Hungary (2021) coincide.
\item[S3] Ecuador (2015) values from Table B.10 of \cite{bacilieri2026firm}, rounded to nearest integer.
\item[S4] From Table 9 of \cite{bacilieri2026firm}.
\item[S5] Midpoint of estimates reported in Table 9 of \cite{bacilieri2026firm}.
\item[S6]  At tail probability $0.01$, Hill estimates align on average with \texttt{igraph} \texttt{plfit}. See Figure \ref{fig:hill_deg_threshold_percentile_against_plfit}.
\item[S7] Nearest $1\%$ from Ecuador (2015) and Hungary (2021), Table C.17 and C.18 of \cite{bacilieri2026firm}.
\item[S8] Ecuador (2015), Table C.16 of \cite{bacilieri2026firm}.
\item[S9] Ecuador (2015), Table C.19 of \cite{bacilieri2026firm}.
\item[S10] Inferred from Ecuador (2015) and Hungary (2021), Table B.11 of \cite{bacilieri2026firm}.
\item[S11] Midpoint between Ecuador (2015) and Hungary (2021), Table B.11 of \cite{bacilieri2026firm}.
\item[S12] Computed from $\mathrm{OLS}(y\sim x)\mathrm{OLS}(x\sim y)=\mathrm{Corr}(x,y)^2$.

\end{tablenotes}
\end{threeparttable}
\caption{Network properties and their target values, based on empirical findings in \cite{bacilieri2026firm}. Denote in-degree by $k^{\mathrm{in}}$, out-degree by $k^{\mathrm{out}}$, in-strength by $s^{\mathrm{in}}$, and out-strength by $s^{\mathrm{out}}$. Mean log-strengths are computed in units of $1$ USD. All correlations, Ordinary Least Squares (OLS), and Total Least Squares (TLS) coefficients are computed on log-transformed variables. Ref. \cite{bacilieri2026firm} report a mean degree of around 30-50, but suggest a 1/3 scaling exponent with size $N$ by studying various countries for which networks are observed over time. We obtain the prefactor 0.86 to obtain a degree of about 40 for $N=10^5$.}
\label{tab:choosing_empirical_values}
\end{table}

%% file: tables/ablation_study.tex
\begin{table}[!htbp]
\centering
\resizebox{\textwidth}{!}{%
\begin{tabular}{@{}>{\raggedleft\arraybackslash}p{6.2cm}rrr@{\hspace{1cm}}rr@{}}
\toprule
 &
  \multicolumn{3}{c}{\multirow{2}{*}{\textbf{for fixed mean $=40$}}} &
  \multicolumn{2}{c}{\textbf{for mean $\bar{k}=0.86N^{1/3}$}} \\
                                                             & \multicolumn{3}{c}{}   & \multicolumn{2}{c}{\textbf{(rounded to nearest $10$)}} \\ \cmidrule(lr){2-4} \cmidrule(l){5-6}
 &
  \textbf{$N=100,000$} &
  \textbf{$N=10,000$} &
  \textbf{$N=1,000$} &
  \textbf{\begin{tabular}[c]{@{}c@{}}$N=10,000$\\ $\bar{k}=20$\end{tabular}} &
  \textbf{\begin{tabular}[c]{@{}c@{}}$N=1,000$\\ $\bar{k}=10$\end{tabular}} \\ \hline
\textbf{Burr with discretization, truncation, and reconciliation} &                 &                &               &                &               \\
mean degree                                                       & 39 (0.917)      & 33.8 (1.48)    & 19.6 (1.54)   & 18.1 (0.933)   & 7.28 (0.804)  \\
out-deg hill exponent                                             & 1.4 (0.0408)    & 1.67 (0.138)   & 5.83 (1.96)   & 1.51 (0.132)   & 2.53 (0.713)  \\
in-deg hill exponent                                              & 2.33 (0.0695)   & 4.57 (0.389)   & 24.7 (6.41)   & 2.9 (0.237)    & 12.9 (3.78)   \\
out-deg log-variance                                              & 2.95 (0.0151)   & 2.92 (0.045)   & 2.51 (0.105)  & 2.43 (0.0437)  & 1.87 (0.111)  \\
in-deg log-variance                                               & 2.01 (0.0118)   & 1.93 (0.028)   & 1.29 (0.0537) & 1.68 (0.0294)  & 1.16 (0.0588) \\
max out-deg                                                       & 30,616 (5,939)  & 3,682 (267)    & 381 (17.4)    & 3,277 (512)    & 330 (50.8)    \\
max in-deg                                                        & 4,007 (561)     & 463 (20.8)     & 73.6 (7.19)   & 450 (29)       & 50 (5.87)     \\ \hline
\textbf{Burr with discretization and truncation}                  &                 &                &               &                &               \\
in-deg mean                                                       & 40 (0.269)      & 39.9 (0.8)     & 35.2 (1.66)   & 20 (0.422)     & 9.84 (0.581)  \\
in-deg hill exponent                                              & 2.31 (0.0687)   & 2.37 (0.217)   & 6.45 (1.95)   & 2.35 (0.227)   & 2.95 (0.884)  \\
in-deg log-variance                                               & 2.02 (0.00805)  & 2.02 (0.0261)  & 1.93 (0.0744) & 1.73 (0.0219)  & 1.4 (0.0578)  \\
max in-deg                                                        & 6,728 (3,722)   & 2,429 (673)    & 380 (18)      & 1,434 (630)    & 247 (69)      \\ \hline
\textbf{Burr with discretization only}                            &                 &                &               &                &               \\
out-deg mean                                                      & 40 (2.17)       & 40 (4.83)      & 40.1 (18)     & 20 (2.42)      & 9.96 (3.43)   \\
in-deg mean                                                       & 40 (0.267)      & 40 (0.871)     & 40 (2.7)      & 20 (0.441)     & 9.97 (0.665)  \\
out-deg hill exponent                                             & 1.39 (0.0416)   & 1.41 (0.132)   & 1.67 (0.553)  & 1.41 (0.131)   & 1.66 (0.538)  \\
in-deg hill exponent                                              & 2.31 (0.0691)   & 2.35 (0.225)   & 2.79 (0.929)  & 2.35 (0.225)   & 2.76 (0.903)  \\
out-deg log-variance                                              & 2.95 (0.015)    & 2.96 (0.0484)  & 2.96 (0.151)  & 2.45 (0.0454)  & 1.99 (0.135)  \\
in-deg log-variance                                               & 2.02 (0.00808)  & 2.02 (0.0256)  & 2.02 (0.0802) & 1.73 (0.022)   & 1.40 (0.0589) \\
max out-deg                                                       & 85,028 (171,357)& 23,628 (34,725)& 6,558 (16,108)& 11,994 (17,927)& 1,603 (2,777) \\
max in-deg                                                        & 6,862 (4,178)   & 3,123 (2,303)  & 1,324 (932)   & 1,525 (998)    & 332 (226)     \\ \hline
\textbf{Burr}                                                     &                 &                &               &                &               \\
out-deg mean                                                      & 40 (3.41)       & 40 (5.13)      & 40 (17.7)     & 20 (2.33)      & 9.97 (3.80)   \\
in-deg mean                                                       & 40 (0.277)      & 40 (0.855)     & 40 (2.74)     & 20 (0.43)      & 9.99 (0.685)  \\
out-deg hill exponent                                             & 1.39 (0.0409)   & 1.4 (0.132)    & 1.59 (0.529)  & 1.41 (0.132)   & 1.6 (0.543)   \\
in-deg hill exponent                                              & 2.31 (0.0681)   & 2.34 (0.221)   & 2.66 (0.911)  & 2.33 (0.226)    & 2.64 (0.91)  \\
out-deg log-variance                                              & 10 (0.0611)     & 10 (0.195)     & 10 (0.612)    & 10 (0.198)     & 10 (0.617)    \\
in-deg log-variance                                               & 3 (0.0182)      & 3 (0.0583)     & 3 (0.183)     & 3 (0.0587)     & 3 (0.185)     \\
max out-deg                                                       & 84,604 (312,946)& 23,879 (38,485)& 6,519 (15,630)& 11,901 (16,318)& 1,581 (3,144) \\
max in-deg                                                        & 6,899 (4,097)   & 3,108 (2,099)  & 1,337 (987)   & 1,563 (1,061)  & 334 (238)     \\ \hline
\end{tabular}%
}
\caption{Ablation study of the effects of discretization, truncation, and reconciliation on degree sequences sampled from the Burr XII distribution $X \sim \mathrm{BurrXII}(c,k,\lambda)$, with tail exponent $\alpha = ck$, log-variance $\nu = \mathrm{Var}(\ln X) = \frac{\frac{\pi^2}{6} + \psi_1(k)}{c^2}$, and mean $\mu = \lambda k\, B\!\left(k - \frac{1}{c},\, 1 + \frac{1}{c}\right)$. Degree sequences are generated by specifying the mean $\mu$, with $\alpha = 3$, $\nu = 3$ for in-degrees and $\alpha = 2$, $\nu = 10$ for out-degrees. 
Each block reports averages and standard deviations (in parentheses) over $5000$ sampled sequences under different combinations of engineering procedures. The table tracks how mean degree, maximum degree, Hill exponents, and log-variances change before and after these procedures, and how the resulting values deviate from their targets across network sizes. Because the out-degree sequence is sampled first and the in-degree sequence is reconciled to match it, reconciliation does not affect out-degree statistics; these are omitted from the ``\textbf{with discretization and truncation}'' block.
Two regimes are considered: a fixed target mean degree $\mu = 40$, and a size-dependent mean $\bar{k} = 0.86 N^{1/3}$ (rounded to the nearest $10$). Without engineering procedures, the sampled sequences exactly match the specified mean $\mu = 40$ and log-variances (3 and 10, respectively), but the estimated Hill exponents differ from the specified values ($\alpha = 3$ and $2$) due to the large log-variances (see Figures \ref{fig:burr_alpha_nu_out_degree_meshgrid}, \ref{fig:burr_alpha_nu_in_degree_meshgrid} and Eq. \eqref{eq:log_log_ccdf_slope}). With discretization, truncation, and reconciliation, systematic distortions emerge, especially in smaller networks: as $N$ decreases, Hill exponents increase and log-variances decrease. These results explain the deviations observed in Figure \ref{fig:stability_across_network_size}, including an upward bias in tail exponents and downward deviation in mean degree for small $N$.}
\label{tab:integerize_truncate_reconcile_degrees_hill001}
\end{table}

%% file: tables/model_architecture_tikz.tex
    \begin{tikzpicture}[
        node distance=1.05cm,
        font=\small,
        every node/.style={transform shape}
    ]

        \node (starting_ingredients)
        [block_losses, text width=6cm, align=center] {%
            Given network topology $\mathbf{A}_{ij}$ with degree sequences
            $\{k_i^{\text{out}}\}_{i=1}^N$, $\{k_j^{\text{in}}\}_{j=1}^N,$
            fitnesses
            $\{f_i^{\text{out}}\}_{i=1}^N,$ 
            $\{f_j^{\text{in}}\}_{j=1}^N,$ and firm-to-industry assignments $\{g_i\}_{i=1}^N$
        };

        \node (weight_matrix_pre_init)
        [block, below=of starting_ingredients, text width=6cm, align=center] {%
            $\mathbf{W}_{ij}^{\text{init}}=$
            $\begin{aligned}
            &(f_i^{\text{out}})^{\theta_{f,\mathrm{out}}} (f_j^{\text{in}})^{\theta_{f,\mathrm{in}}}
            (k_i^{\text{out}})^{\theta_{k,\mathrm{out}}} (k_j^{\text{in}})^{\theta_{k,\mathrm{in}}} A_{ij}
            \end{aligned}$
        };

        \node (weight_matrix)
        [block, below=of weight_matrix_pre_init, yshift=-8mm]
        {Weight matrix $\mathbf{W}_{ij}$};

        \node (adj_matrix)
        [block, left=of weight_matrix, text width = 2.5cm,xshift=-3cm]
        {Adjacency matrix $\mathbf{A}_{ij}$};

        \node (deg_seq)
        [block, below=of adj_matrix, yshift=-2mm]
        {In, out-degree sequences};

        \node (str_seq)
        [block, below=of weight_matrix, yshift=-2mm]
        {In, out-strength sequences};

        \node (str_losses)
        [block_losses, below=of str_seq, yshift=-3mm] {
            \begin{enumerate}[topsep=0pt, itemsep=0pt]
                \setcounter{enumi}{2}
                \item Variance of log strengths
                \item Hill exponent strengths
                \item TLS$(\text{in-str},\text{out-str})$
            \end{enumerate}
        };

        \node (joint_distri_losses)
        [
            block_losses,
            align=left,
            text width = 3.75cm,
            inner sep=2.5pt,
            minimum width=0pt,
            minimum height=0pt,
            left=of str_losses,
            xshift=-0.1cm
        ] {
            \begin{enumerate}[topsep=0pt, itemsep=0pt]
                \setcounter{enumi}{0}
                \item Correlations
                \item OLS coefficients
            \end{enumerate}
        };

        \node (total_loss)
        [block, below=of str_losses, yshift=-4mm]
        {$\text{Loss}_{\text{optuna}}$};

        \draw[arrow]
            (starting_ingredients.south)
            -- node[midway, fill=white, inner sep=1pt] (param_node) {$\theta_{k,\mathrm{in}} \;,\;\theta_{k,\mathrm{out}}\;,\;\theta_{f,\mathrm{in}}\;,\;\theta_{f,\mathrm{out}}$}
            (weight_matrix_pre_init.north);

        \draw[arrow]
        (weight_matrix_pre_init) to
        node[
            pos=0.55,
            anchor=east,        
            xshift=-1.8mm,      
            inner sep=0pt,
            outer sep=0pt,
            align=right,
            text width=5.6cm
        ] {%
        $\displaystyle
        \frac{\mathrm{IOT}_{g_i g_j}}
        {\sum\limits_{m\in g_i}\sum\limits_{n\in g_j} \mathbf{W}_{mn}^{\text{init}}}
        \, \mathbf{W}_{ij}^{\text{init}}$}
        (weight_matrix);

        \draw[arrow] (weight_matrix.west) -- node[below] {$\mathbbm{1}(\mathbf{W}_{ij}>0)$} (adj_matrix.east);
        \draw[arrow] (adj_matrix) -- node[anchor=east] {\shortstack{Row sums \& \\ column sums}} (deg_seq);
        \draw[arrow] (weight_matrix) -- node[anchor=east] {\shortstack{Row sums \& \\ column sums}} (str_seq);

        \draw[arrow] (str_seq) -- (str_losses);
        \draw[arrow] (deg_seq.south) -- (joint_distri_losses.north);
        \draw[arrow] (str_seq.south) -- (joint_distri_losses.north);
        \draw[arrow] (joint_distri_losses.south) -- (total_loss.west);

        \draw[arrow] (str_losses.south) -- (total_loss.north);

        \draw[arrow]
            (total_loss.south) -- ++(0,-3mm) -- ++(3.5cm,0)
            |- node[pos=0.25, rotate=270, anchor=south]{Optuna}
            (param_node.east);

    \end{tikzpicture}%

%% file: tables/mean_std_of_best_theta_from_optuna_search.tex
\begin{table}[!htbp]
\centering

\begin{tabular}{@{}lcc@{}}
\toprule
Parameter & Mean & \multicolumn{1}{l}{Standard deviation} \\
\midrule
$\theta_{k,\mathrm{in}}$ & $-0.0265$ & $0.0757$ \\
$\theta_{k,\mathrm{out}}$ & $-0.697$ & $0.0644$ \\
$\theta_{f,\mathrm{in}}$ & $-1.35$, $1.37$ & $0.11$, $0.12$ \\
$\theta_{f,\mathrm{out}}$ & $-1.27$, $1.29$ & $0.0756$, $0.0623$ \\
\bottomrule
\end{tabular}

\caption{Mean and standard deviation of the optimal parameters obtained from Optuna. The statistics are computed from $200$ independent runs, each experiment consisting of $200$ trials. The parameter $\theta_{k,\mathrm{in}}$ is centered around $0$, while $\theta_{k,\mathrm{out}}$ is consistently negative. In contrast, $\theta_{f,\mathrm{in}}$ and $\theta_{f,\mathrm{out}}$ are either both negative or positive (see Figure \ref{fig:optimal_parameters_from_optuna}), so we report both cases separately.}
\label{tab:optima_parameters_from_optuna_stats}
\end{table}

%% file: tables/100_networks_using_pos_neg_roundedneg_thetas.tex
\begin{table}[!htbp]

\setlength{\heavyrulewidth}{1.3pt}
\setlength{\lightrulewidth}{0.8pt}
\setlength{\cmidrulewidth}{0.8pt}

\centering
\resizebox{0.88\textwidth}{!}{%
\begin{tabular}{@{}lrrrr@{}}
\toprule
Property & Positive & Negative & Rounded negative & Empirical \\
\midrule
$\theta_{k,\mathrm{in}}$ & -0.0265 & -0.0265 & 0 & \\
$\theta_{k,\mathrm{out}}$ & -0.697 & -0.697 & -0.7 & \\
$\theta_{f,\mathrm{in}}$ & 1.37 & -1.35 & -1.3 & \\
$\theta_{f,\mathrm{out}}$ & 1.29 & -1.27 & -1.3 & \\
\midrule
Number of nodes & 100,000 (0) & 100,000 (0) & 100,000 (0) & 100,000 \\
Number of edges & 3,855,343 (23,605) & 3,857,177 (24,644) & 3,862,201 (26,739) & 4,000,000 \\
Share of supplier-only firms & 0 (0) & 0 (0) & 0 (0) &  \\
Share of customer-only firms & 26.3 (0.139) & 26.3 (0.16) & 26.3 (0.146) &  \\
Mean degree & 38.6 (0.236) & 38.6 (0.246) & 38.6 (0.267) & 40 \\
\midrule
Max $k^{\text{in}}$ & 2,787 (360) & 2,823 (294) & 2,847 (328) &  \\
Max $k^{\text{out}}$ & 21,256 (1,837) & 20,786 (1,994) & 21,076 (2,160) &  \\
Mean log $k^{\text{in}}$ & 3.28 (0.00911) & 3.28 (0.0088) & 3.28 (0.00795) & 3 \\
Var log $k^{\text{in}}$ & 1.31 (0.0082) & 1.31 (0.00859) & 1.31 (0.00826) & 2 \\
Mean log $k^{\text{out}}$ & 2.19 (0.00621) & 2.19 (0.00664) & 2.19 (0.00654) & 2 \\
Var log $k^{\text{out}}$ & 2.96 (0.0111) & 2.96 (0.0125) & 2.97 (0.0124) & 3 \\
\midrule
Mean log $s^{\text{in}}$ & 10.8 (0.0349) & 10.8 (0.0424) & 10.9 (0.0366) & 10 \\
Var log $s^{\text{in}}$ & 9.14 (0.102) & 8.96 (0.0973) & 8.9 (0.11) & 9 \\
Mean log $s^{\text{out}}$ & 10.8 (0.0399) & 10.9 (0.0392) & 10.9 (0.0405) & 10 \\
Var log $s^{\text{out}}$ & 9.51 (0.0901) & 9.25 (0.0749) & 9.25 (0.0939) & 8 \\
\midrule
LWCC & 95,517 (71) & 95,525 (61) & 95,526 (75) &  \\
Avg path length & 2.74 (0.0234) & 2.75 (0.0186) & 2.75 (0.0214) & 3 \\
Degree assortativity & $-0.0679 (0.00379)$ & $-0.0668 (0.00363)$ & $-0.0668 (0.00444)$ & $[-0.2,-0.01] < 0$ \\
Reciprocity & 0.00647 (0.000608) & 0.00655 (0.000555) & 0.00656 (0.000554) & [0.03, 0.05] \\
Avg clustering & 0.0946 (0.00603) & 0.0929 (0.00563) & 0.0935 (0.0068) & [0.19, 0.28] \\
Global clustering & 0.0217 (0.000777) & 0.0218 (0.000707) & 0.0218 (0.000868) & low \\
\midrule
Hill $\alpha$, $k^{\text{in}}$ & 2.55 (0.0321) & 2.54 (0.0328) & 2.54 (0.0306) & 2.5 \\
Hill $\alpha$, $k^{\text{out}}$ & 1.46 (0.0144) & 1.47 (0.0177) & 1.47 (0.0162) & 1.5 \\
Hill $\alpha$, $s^{\text{in}}$ & 1.06 (0.0215) & 1.07 (0.0223) & 1.08 (0.0193) & 1 \\
Hill $\alpha$, $s^{\text{out}}$ & 1.07 (0.0201) & 1.08 (0.0229) & 1.07 (0.0199) & 1 \\
Hill $\alpha$, influence & 1.45 (0.0223) & 1.46 (0.0224) & 1.44 (0.0219) & [1.2, 1.3] \\
Hill $\alpha$, weights & 1.06 (0.0149) & 1.07 (0.016) & 1.07 (0.0154) & [1.1, 1.2] \\
\midrule
Corr: $k^{\text{in}}\sim k^{\text{out}}$ & 0.551 (0.00296) & 0.551 (0.00294) & 0.551 (0.00281) & 0.55 \\
Corr: $s^{\text{in}}\sim s^{\text{out}}$ & 0.508 (0.00485) & 0.506 (0.0044) & 0.502 (0.0051) & 0.5 \\
Corr: $s^{\text{out}}\sim k^{\text{out}}$ & 0.453 (0.00381) & 0.455 (0.00311) & 0.44 (0.00443) & 0.5 \\
Corr: $s^{\text{in}}\sim k^{\text{in}}$ & 0.565 (0.00352) & 0.567 (0.00332) & 0.584 (0.0028) & 0.75 \\
\midrule
OLS: $s^{\text{out}}\sim k^{\text{out}}$ & 0.812 (0.00947) & 0.803 (0.00746) & 0.777 (0.0101) & 0.76 \\
OLS: $k^{\text{out}}\sim s^{\text{out}}$ & 0.253 (0.00171) & 0.257 (0.00164) & 0.249 (0.00218) & 0.33 \\
OLS: $s^{\text{in}}\sim k^{\text{in}}$ & 1.49 (0.0112) & 1.48 (0.011) & 1.52 (0.0122) & 1.4 \\
OLS: $k^{\text{in}}\sim s^{\text{in}}$ & 0.214 (0.00207) & 0.217 (0.00165) & 0.224 (0.00164) & 0.4 \\
\midrule
TLS: $s^{\text{in}}\sim s^{\text{out}}$ & 0.962 (0.00945) & 0.969 (0.00995) & 0.962 (0.0127) & 1 \\
TLS: $k^{\text{in}}\sim k^{\text{out}}$ & 0.494 (0.0036) & 0.494 (0.00373) & 0.494 (0.00339) & 0.7 \\
\midrule
IOT RMSE (100 MUSD) & 0.00134 (0.000947) & 0.00137 (0.000818) & 0.00138 (0.00085) &  \\
\bottomrule
\end{tabular}%
}
\caption{Properties of the networks generated under different parameter settings $( \theta_{k,\mathrm{in}} \;,\;\theta_{k,\mathrm{out}}\;,\;\theta_{f,\mathrm{in}}\;,\;\theta_{f,\mathrm{out}} )$, based on $100$ networks of $100,000$ firms that match the $2015$ Hungarian input-output table. Entries report means, with standard deviations in parentheses. ``Positive'' and ``Negative'' denote parameter settings with positive and negative fitness respectively, while ``Rounded negative'' uses rounded values of the negative setting. In-degree is the number of suppliers, out-degree is the number of customers. 
Supplier-only firms sell to at least one other firm but buy from none $(k^{\mathrm{in}}=0)$, whereas customer-only firms buy from at least one other firm but sell to none $(k^{\mathrm{out}}=0)$. 
LWCC stands for Largest Weakly Connected Component. Mean of both log in- and out-strengths are computed in units of $1$ USD. OLS and TLS stand for Ordinary and Total Least Squares, and $\alpha$ stands for a tail exponent computed using a Hill estimator. All values are reported to three significant figures, except for the number of nodes, edges, and LWCC, which are reported as integers. All three parameter settings closely match the empirical targets with low variability. 
The $R^2$ coefficient is obtained from $R^2 = \mathrm{Corr}(y,x)^2$ for $\mathrm{OLS}{:y \sim x}$. The RMSE of the aggregated matrix is negligible.}
\label{tab:results_stability}
\end{table}